\documentclass[graybox]{svmult}
\usepackage{mathrsfs}

\usepackage{mathptmx}       
\usepackage{helvet}         
\usepackage{courier}
\usepackage{makeidx}         
\usepackage{graphicx}        
\usepackage{caption}
\usepackage{multicol}        
\usepackage[bottom]{footmisc}
\usepackage{amsmath,amssymb,amsfonts}        
\makeindex             

\begin{document}
\title*{Asian Option Pricing via Laguerre Quadrature: A Diffusion Kernel Approach}
\author{P. G. Morrison}
\institute{Peter Morrison \at University of Technology, Sydney, Broadway 2000, \email{peter.morrison@uts.edu.au}
}
%
%
\maketitle
\abstract{
This paper will demonstrate some new techniques for developing the theory of Asian (arithmetic average) options pricing. We discuss the basic derivation of the diffusion equations, and how various techniques from potential theory can be applied to solve these complex expressions. The Whittaker-type confluent hypergeometric functions are introduced, and we discuss how these functions are related to other systems including Mehler-Fock and modified Bessel functions. We close with a brief analysis of some index transforms and the kernels related to these integral transforms.
}

\section{Introduction}
\label{sec:1}
Asian (arithmetic average) options are a derivative instrument written on the difference between the time average of the stock price over the term, and some nominal strike. Floating strike options have payoff given by the max/min of the difference between the time average and the stock price on the execution date. These derivatives have useful properties when offsetting risk for continuously produced goods such as commodities and electricity. Other use cases could be in order to reward executive performance above some average benchmark, say the average stock price is greater than some nominal value over the term of the option. It is harder to influence in a negative way vs standard European contracts; indeed it is well known that stock volatility increases towards the execution date of European options which leads to inverse incentives, i.e. increases the probability of stock price pumps by executives near the time of expiry.

If one were to think in terms of peak average performance, rather than the performance on any particular one day, it is obvious that such contracts as we shall describe in this paper have great utility in this area. For a commodities producer, the value in such contracts is that it enables producers to lock in prices for known deposits and stabilise pricing outcomes separate to the spot market, given known fixed production costs at any given time. A minerals producer might for example know that they can produce at an average rate 10,000 tons of ore at a fixed concentration of gold per day. Pricing an option on the average amount of gold produced over a month then becomes of utility when locking in a price for the total amount of gold produced, which in turn can be used to stabilise the costs to market the commodity by using this sort of technique. Indeed, this paper concerns the deep problem of the nature of the average of an exponential Brownian motion. 

There is a deep, known link between the diffusion problems we associate with options pricing problems and diffusion theory. It is known that the Black-Scholes equation may be transformed to the heat equation, other results include Feynman-Kac for generalised potentials. Other exotic options, e.g. in CDS pricing equation, one may obtain the diffusion equation related to the CIR process, which has probability function given by a non-central chi-squared distribution. As shall be demonstrated in the following calculations, there is a fundamental relationship between diffusion systems, special functions and spectral theory which is of great use in understanding these types of problems in the theory of stochastic processes.

\section{Review}

The work contained within this paper pays homage to the pioneers in
the field of spectral analysis, principal amongst them Feller \cite{feller1952parabolic}, who explored the methods of semi-groups and partial differential
equations in order to understand stochastic processes and other random
variables. These initial researches, carried out in the 1950s, can
be seen as the ancestor of the theories of stochastic analysis, especially
PDE methods. However, it is a lesser-known topic, with the applications
of these types of kernel analysis often obscured by the bright lights
of martingales and stochastic calculus. The construction of kernels
to solve differential equations is an older question again, and although
it is well-known in the fields of physics that an understanding of
kernel solutions can be gained through spectral decomposition, this
observation has escaped much detailed investigation in the field of
quantitative finance. 

Spectral decomposition is an elegant way of describing the transition
probability density of any system with a spectrum, and the linkage
between resolvent or Green's function methods and the probability
density of a diffusion is given through Laplace transformation. We
note the papers of Albanese et. al \cite{albanese2007laplace,albanese2007transformations} examine this problem in general, and have found a number of systems with spectral
decompositions that are of finite type. On a deeper level, underlying
this whole infrastructure is the effort of Mehler \cite{mehler1866ueber},
as expressed and popularised through the Feynman kernel, path integral
method c. 1948 \cite{feynman2010quantum}. This formula is a compact description
of a kernel using a set of base states, which in this case is given
by the Hermite polynomials. The search for other such formulae is
an active area of investigation and we shall touch on some simple
topics related to this historical example. As an example of a system
related to our investigation, use of the various types of special
functions as a result of path integration in the hyperbolic plane
may be found in \cite{grosche1988path}. The authors in this case have succeeded
in deriving many formulae that appear in a different context to our
research. In many ways, the idea of hyperbolic diffusion is closely
allied to the types of hyperbolic random variables that underpin the
diffusion problems in this work. Further tabulations of path integrals
may be found in \cite{grosche1998handbook}, in particular the results concerning
the resolvent of Whittaker functions and the Morse potential.

The link between special function theory and arithmetic average or
Asian option pricing has been explored by a number of authors. The
foundational paper in the field is taken to be Geman and Yor \cite{geman1993bessel}, who established the connection between the theory of square
Bessel processes, stochastic calculus, and exponential Brownian motion.
The topic was explored from a different perspective by Linetsky \cite{linetsky2004spectral}, who has demonstrated that spectral theory may be used
to calculate Asian options prices. Both of these papers present formidable
work, and our research has shown that many of the results which are
gained through sheer force of will and brute mathematical force in
these works can be found through simpler means, via the techniques
of spectral decomposition of the kernel. The stochastic calculus in
\cite{geman1993bessel} shows that prima facie the concept of exponential Brownian
motion is the important branch of mathematics on which to grasp; one
is presented with the idiom of time-averaged exponential Brownian
motion. The functional analysis of these types of systems is carried
out in detail by Pintoux and Privault \cite{pintoux2010direct,pintoux2011dothan}, where the
authors were able to obtain some exact solutions to the Asian option
pricing problem by deriving some spectral decompositions of polynomial
initial conditions. In particular, the papers in \cite{pintoux2010direct,pintoux2011dothan}
rely on many combinatorical identities and are complicated to verify
and replicate. Our methods represent a parallel line of thought which
arrives at similar results by a completely different path. A simpler
approach has been illustrated through the works of Sousa and Yakubovich
\cite{sousa2017spectral}, who have shown that by understanding the nature of
kernel solutions, eigenfunctions and integral transforms one may gain
many of the advantages of the PDE method while avoiding necessary
complication in the matter. Further, Craddock \cite{craddock2014integral} has
shown that analysis of the fundamental solution for Asian options
pricing can be understood through the use of special functions and
the sine transform. 

We shall expand upon this method in this paper and bring this program
to fruition by a process of synthesis of the results from these disparate
groups. Our aim is to find direct ways of calculating Asian options
prices that will be of use to any trader in the same way that European
options prices may be found via the Black-Scholes equation in the
simplest of cases. Such formulae have proved elusive in the past,
and the notational complexity of the field makes our job much more
difficult. By stripping away this additional complication, we shall
show that the answers to many of the problems involved in pricing
these types of exotic options reduce to a small number of simply addressed
questions in integration theory.

We must note here the use of special functions known as the Bessel
polynomials in this paper. These polynomials have many properties
peculiar to the problems in diffusion we shall consider, and it can
be hard to find established or tabulated results that one can expect
to find for more well-known groups of orthogonal polynomials. The
interested reader is directed to the works of Krall and Frink \cite{krall1949new}, who along with Grosswald \cite{grosswald1951some} wrote the pioneering
works on the topic, see also \cite{hamza1974properties} for further information
and references on integration theory of Bessel polynomials. These
polynomials, familiar to filter design and signal processing, are
less known in modern mathematics. Their many useful properties in
describing systems with discrete and continuous spectral components
shall be one principal focus explored in this calculation. 

Examples of index transforms and related integrals for Whittaker,
modified Bessel and Mehler-Fock functions may be found in the tabulated
works in \cite{gradshteyn2014table, prudnikov1986special,prudnikov1988integrals,bateman1954tables, oberhettinger1961tables,oberhettinger2012tables}. The monumental
works of Oberhettinger \cite{oberhettinger1961tables,oberhettinger2012tables} are of great use in the following
sections. Unfortunately, like all sets of integral tables, deficiencies
abound, and part of this research is devoted to unpicking the methods
by which the known and established results in index transform theory
may be derived.

\section{Notation and Mathematical Details}

This paper shall be primarily concerned with various sets of special
functions which we shall now define. There are other notational necessities
that must also be set up for the calculation proper. We shall not
be focused in this work on the exact nature of how these functions
arise, for our purpose it is sufficient to know that they exist and
are useful to our aim of analysing diffusion processes of a type that
is central to our work.

\subsection{Special functions}

The important special functions used in this paper are defined as
follows:

\subsubsection{Whittaker Functions}

The Whittaker function, denoted $W_{\alpha,\beta}(z)$, is a type
of confluent hypergeometric function. It is the solution of the differential
equation:

\begin{equation}
\dfrac{d^{2}f}{dz^{2}}+\left(-\dfrac{1}{4}+\dfrac{\alpha}{z}+\dfrac{1/4-\beta^{2}}{z^{2}}\right)f=0
\end{equation}

The indices of the function may take any value in the complex plane,
however it is important to realise that this function is multi-branched
and care must be taken to choose appropriate sets of parameters. For
our purposes, as we shall demonstrate, the positive plane defines
a continuous solution which may be used to build a set of eigenfunctions.
The second solution for the Whittaker differential equation is not
used in this proof. Applications of the Whittaker function include
the use in the index Whittaker transform \cite{sousa2017spectral} where, in
a similar way to how functions are mapped to other functions via a
Laplace or Fourier transform, in this case we integrate over the index
of the function via:

\begin{equation}
F(x)=\dfrac{1}{\pi^{2}x^{2}}\int_{0}^{\infty}|\Gamma(-\mu+1/2+ip)|^{2}W_{\mu,ip}(x)f(p)p\sinh(2\pi p)dp
\end{equation}

This is the most generalised form of the index transformation. Note
that this is of half-index in the gamma function. An inversion integral
of the form:

\begin{equation}
f(p)=\int_{0}^{\infty}W_{\mu,ip}(x)F(x)dx
\end{equation}
is known to exist. We shall rely on the relationships between the
confluent hypergeometric function and other special functions to compute
a number of different integrals related to the solution of various
PDEs. Note the useful hyperbolic identity:

\begin{equation}
\dfrac{\sinh(2\pi p)}{\cosh(\pi p)}=2\sinh(\pi p)
\end{equation}

\subsubsection{Modified Bessel Functions}

The modified Bessel function $K_{\nu}(z)$ is another type of hypergeometric
function. One may relate it to the restriction of the Whittaker function
via:

\begin{equation}
W_{0,\nu}(2z)=\sqrt{\dfrac{2z}{\pi}}K_{\nu}(z)
\end{equation}
The modified Bessel function therefore satisfies the differential
equation defined through:
\begin{equation}
\dfrac{d^{2}f}{dz^{2}}+\left(-\dfrac{1}{4}+\dfrac{1/4-\nu^{2}}{z^{2}}\right)f=0
\end{equation}
Using the scaling $x=z/2$, we may transform the differential equation,
so in the new variables we have:

\begin{equation}
\dfrac{1}{4}f_{xx}+\left(-\dfrac{1}{4}+\dfrac{1/4-\nu^{2}}{4x^{2}}\right)f=0
\end{equation}
or
\begin{equation}
f_{xx}+\left(-1+\dfrac{1/4-\nu^{2}}{x^{2}}\right)f=0
\end{equation}
Substituting $f(x)=\sqrt{x}U(x)$, we find the modified Bessel differential
equation:
\begin{equation}
x^{2}U_{xx}+xU_{x}-(\nu^{2}+x^{2})U(x)=0
\end{equation}

This is an important type of hypergeometric function as shown in the
calculation. In an analogous way to the index Whittaker transform,
one may define an index transform for the modified Bessel function
known as the Kontorovich-Lebedev transform pair. Substituting using
the functional relationship, we have:
\begin{equation}
f(p)=\int_{0}^{\infty}W_{0,ip}(2x)F(x)dx
\end{equation}
\begin{equation}
F(x)=\dfrac{1}{\pi^{2}x^{2}}\int_{0}^{\infty}|\Gamma(1/2+ip)|^{2}W_{0,ip}(2x)f(p)p\sinh(2\pi p)dp
\end{equation}

and, using the gamma function identity
\begin{equation}
|\Gamma(ip+1/2)|^{2}=\dfrac{\pi}{\cosh(\pi p)}
\end{equation}

we find:
\begin{equation}
W_{0,\nu}(2z)=\sqrt{\dfrac{2z}{\pi}}K_{\nu}(z)
\end{equation}
\begin{equation}
f(p)=\int_{0}^{\infty}\sqrt{\dfrac{2x}{\pi}}K_{ip}(x)F(x)dx=\sqrt{\dfrac{2}{\pi}}\int_{0}^{\infty}\sqrt{x}K_{ip}(x)F(x)dx
\end{equation}
\begin{equation}
F(x)=\dfrac{1}{\pi^{2}x^{2}}\int_{0}^{\infty}\dfrac{\pi p\sinh(\pi p)}{\cosh(\pi p)}\sqrt{\dfrac{2x}{\pi}}K_{ip}(x)f(p)dp
\end{equation}
\begin{equation}
=\sqrt{\dfrac{2x}{\pi}}\dfrac{2}{\pi x^{2}}\int_{0}^{\infty}p\sinh(\pi p)K_{ip}(x)f(p)dp
\end{equation}

As the same numerical factor appears on both sides of the transform
and inverse it may be moved to either side, similarly with the square
root factor, and we may write:
\begin{equation}
f(p)=\int_{0}^{\infty}K_{ip}(x)F(x)dx
\end{equation}
\begin{equation}
F(x)=\dfrac{2}{\pi^{2}x}\int_{0}^{\infty}K_{ip}(x)f(p)p\sinh(\pi p)dp
\end{equation}

This transform pair is known as the Kontorovich-Lebedev transform.
It is a crucial part of our calculation method, and a number of our
results depend entirely on its computation using some clever tricks.
Note that the integration in the second part of the transform pair
is over the index of the function, which requires a lot more deft
and tact than integrating over the functional variable. This is a
delicate matter requiring some involved mathematics which we shall
encounter many times throughout this work.

\subsubsection{Associated Legendre Functions}

The final special function we shall require relates to both the Kontorovich-Lebedev
and Whittaker transforms in the following way. If we look at the differential
equation for the modified Bessel function at half-integral values:

\begin{equation}
x^{2}f_{xx}+2(1-\alpha)xf_{x}+(\alpha^{2}-\alpha+1/4-\dfrac{1}{4}(1+2k)^{2}-x^{2})f(x)=0
\end{equation}
which we write as $\hat{\mathcal{H}}f(x)=0$. The solution of this
differential equation is given by:
\begin{equation}
f(x)=Cx^{\alpha-1/2}K_{k+1/2}(x)
\end{equation}
where we neglect the second part of the solution due to boundary conditions.
The Laplace transform, which we write as
\begin{equation}
\mathscr{F}\left[f(x)\right]=F(u)
\end{equation}
transforms the differential equation such that we have:
\begin{equation}
\mathscr{F}\left[\hat{\mathcal{H}}f(x)\right]=0
\end{equation}
where the new differential equation is defined through
\begin{equation}
(u^{2}-1)\dfrac{\partial^{2}F}{\partial u^{2}}+2u(\alpha+1)\dfrac{\partial F}{\partial u}+(\alpha^{2}+\alpha-k(k+1))F(u)=0
\end{equation}
and we have neglected boundary terms due to the initial condition,
which is permissible for the purposes of this crude analysis. The
solution to this differential equation, which is related to spherical
harmonic functions, is given by the generalised Legendre function
defined by:
\begin{equation}
F(u)=(u^{2}-1)^{-\alpha/2}C\mathcal{P}_{k}^{\alpha}(u)
\end{equation}

Again, we neglect the second solution to the differential equation
as it is not useful in the calculation for our boundary conditions.
The function $\mathcal{P}_{ip-1/2}^{\alpha}(u)$, i.e. where we have
the special values $k(k+1)=-p^{2}+1/4$ is the Mehler-Fock or conical
functions and is well known in many applications in physics. It is
important to understand that we are taking the function to be defined
for the range $[1^{+},\infty)$. In an analogous way, the Mehler-Fock
transform may be derived from the Kontorovich-Lebedev transform pair
by using the following integral relationship \cite{gradshteyn2014table}:
\begin{equation}
\int_{1}^{\infty}e^{-xv}(v^{2}-1)^{-\mu/2}\mathcal{P}_{ip-1/2}^{\mu}(v)dv=\sqrt{\dfrac{2}{\pi}}x^{\mu-1/2}K_{ip}(x)
\end{equation}
This represents the result obtained prior through the Laplace transform
of the differential equations. If we then take the Kontorovich-Lebedev
transform, we will have:

\begin{equation}
f(p)=\int_{0}^{\infty}K_{ip}(x)F(x)dx
\end{equation}

\begin{equation}
=\int_{0}^{\infty}F(x)dx\sqrt{\dfrac{\pi}{2}}x^{-\mu+1/2}\int_{1}^{\infty}e^{-xv}(v^{2}-1)^{-\mu/2}\mathcal{P}_{ip-1/2}^{\mu}(v)dv
\end{equation}
\begin{equation}
=\sqrt{\dfrac{\pi}{2}}\int_{1}^{\infty}dv.\mathcal{P}_{ip-1/2}^{\mu}(v)(v^{2}-1)^{-\mu/2}\int_{0}^{\infty}dx.F(x)x^{-\mu+1/2}e^{-xv}
\end{equation}

Denoting
\begin{equation}
S_{1}(v)=(v^{2}-1)^{-\mu/2}\int_{0}^{\infty}dx.F(x)x^{-\mu+1/2}e^{-xv}    
\end{equation}

which is a Mellin-Laplace composition, we have therefore the forward
transform:
\begin{equation}
\sqrt{\dfrac{\pi}{2}}\int_{1}^{\infty}dv.\mathcal{P}_{ip-1/2}^{\mu}(v)S_{1}(v)=f(p)
\end{equation}

On the other hand, the inverse transformation may be easily found
using the same method, and we obtain:
\begin{equation}
F(x)=\dfrac{2}{\pi^{2}x}\int_{0}^{\infty}K_{ip}(x)f(p)p\sinh(\pi p)dp
\end{equation}
\begin{equation}
K_{ip}(x)=\sqrt{\dfrac{\pi}{2}}x^{-\mu+1/2}\int_{1}^{\infty}e^{-xv}(v^{2}-1)^{-\mu/2}\mathcal{P}_{ip-1/2}^{\mu}(v)dv
\end{equation}

\begin{equation}
F(x)=\dfrac{2}{\pi^{2}x}\int_{0}^{\infty}K_{ip}(x)f(p)p\sinh(\pi p)dp
\end{equation}

\begin{equation}
=\dfrac{2}{\pi^{2}x}\int_{0}^{\infty}dp.\sqrt{\dfrac{\pi}{2}}x^{-\mu+1/2}\int_{1}^{\infty}dv.e^{-xv}(v^{2}-1)^{-\mu/2}\mathcal{P}_{ip-1/2}^{\mu}(v)f(p)p\sinh(\pi p)
\end{equation}
\begin{equation}
=\dfrac{2x^{-\mu-1/2}}{\pi^{2}}\sqrt{\dfrac{\pi}{2}}\int_{1}^{\infty}dv.e^{-xv}(v^{2}-1)^{-\mu/2}\int_{0}^{\infty}dp.\mathcal{P}_{ip-1/2}^{\mu}(v)p\sinh(\pi p)f(p)
\end{equation}
\begin{equation}
F(x)=\int_{1}^{\infty}dv.e^{-xv}(v^{2}-1)^{-\mu/2}S_{2}(v)
\end{equation}
The functional equation appearing in the inner integral is the Mehler-Fock
transformation. The standard choice of paired, normalised transforms
is given by:
\begin{equation}
F(v)=\dfrac{1}{\pi}\int_{0}^{\infty}dp.|\Gamma(1/2-\mu+ip)|\mathcal{P}_{ip-1/2}^{\mu}(v)p\sinh(\pi p)f(p)
\end{equation}
with inverse
\begin{equation}
f(p)=\int_{1}^{\infty}dv.\mathcal{P}_{ip-1/2}^{\mu}(v)F(v)
\end{equation}

Although these formulae appear forbidding, as we shall show, there
is a commonality of structure between these different groups of special
functions which renders many problems solvable in similar ways. Tables
of these transforms are available in some instances, for example \cite{gradshteyn2014table, oberhettinger1961tables,oberhettinger2012tables, prudnikov1986special,prudnikov1988integrals}, however, there is little to no description
of a general method for finding these formulae. This paper shall outline
in detail how one goes about deriving these necessary integrals for
the solution of various problems in diffusion.

\subsubsection{Parabolic Cylinder and Hermite Functions}

The final type of special functions necessary for understanding the
details of the calculations contained herein is the parabolic cylinder
function. These are given by the special case of the Whittaker function
via the connection formula:

\begin{equation}
W_{\mu+1/4,1/4}\left(x\right)=D_{2\mu}\left(\sqrt{2x}\right)\dfrac{x^{1/4}}{2^{\mu}}
\end{equation}
Parabolic cylinder functions satisfy the Weber differential equation:

\begin{equation}
f_{zz}+\left(\nu+\dfrac{1}{2}-\dfrac{z^{2}}{4}\right)f(z)=0
\end{equation}
\begin{equation}
f(z)=cD_{\nu}(z)
\end{equation}
The parabolic cylinder functions and the Hermite polynomials are linked
by the formula:
\begin{equation}
D_{n}(\sqrt{2}x)=2^{-\nu/2}e^{-x^{2}/2}H_{n}(x)
\end{equation}
Indeed, the common factor in all these groups of special functions
is that they all may be described as a limiting case of the confluent
hypergeometric function. 

\subsubsection{Other Sundries}

We shall briefly define all the other necessary objects for the calculation
in this paper. Let us begin with the representation of scalars. A
single valued object (number) without functional dependence shall
be denoted $\lambda$. A set of eigenvalues with indicial parameter
$n$ shall be given the symbol $\lambda_{n}$, and Latin $n$ represents
a discrete set as opposed to a continuous set defined through e.g.
$\lambda_{u}$. Functional dependence will be illustrated through
the use of both discrete $\psi_{n}(x)$ and continuous $\Phi_{u}(x)$.

The kernel solution of a diffusion equation shall be written as $\mathcal{K}(x,y;t)$
for the case of two positions and one time. We indicate that there
is a difference between the time and space variables by use of the
comma. Wherever possible an effort has been made to ensure that the
notation used is consistent for each application. The central focus
of the paper is on the calculation of these types of kernels in an
effort to understand the nature of the dynamical solutions to the
diffusion equations.

We shall make a certain definition in terms of style of writing nested
integrals. This paper shall encounter many of these types of formulae
and they quickly become entangled without a solid start. We make the
operational statement that all integral variables are to be written
to the left and associated with their limit. For example, the integral:

\begin{equation}
I=\int_{0}^{\infty}dx.f(x)\times\int_{1}^{\infty}dy.g(x,y)
\end{equation}
is to be read as the integral of $g(x,y)$ over $y$, then the result
of this times $f(x)$ integrated over $x$. Many of the calculations
in this work stem from the use of the Fubini-Tonelli theorem and the
integrals can become lengthy. For this reason, this convention is
one we follow out of necessity so as to avoid confusion.

Finally, we must define the idea of a transform. This paper deals
with the Whittaker, Kontorovich-Lebedev and Mehler-Fock transforms
in depth, and also obliquely touches upon the Laplace, Fourier, cosine,
Mellin and sine transforms. In the interests of book-keeping, we have
the following approach to offer. When a function is transformed, if
it is a single parameter function, we have automatically no confusion
if we write:

\begin{equation}
F(p)=\mathscr{F}\left[f(x)\right]
\end{equation}
and, under the axiom of invertability we have:
\begin{equation}
f(x)=\mathscr{F}^{-1}\left[F(p)\right]
\end{equation}

We shall always use the cursive script in this fashion when referring
to a transformation of some type. If we have a function with multiple
parameter, such as $\mathcal{K}(x,y)=\mathcal{K}(x,y;0)$, it is necessary
to specify the co-ordinate over which the transform is occurring.
Often the context will make the integration variable in the transform
superfluous and we shall omit it for brevity and refer to the type
of transform by name, or explicitly define the integrals whenever
possible.

\section{Basic Diffusion Theory}
\label{sec:2}
We may write a generic diffusion equation in the form:

\begin{equation}-\dfrac{\partial u}{\partial\tau}=\dfrac{\sigma^{2}}{2}\dfrac{\partial^{2}u}{\partial x^{2}}+\mu\dfrac{\partial u}{\partial x}+V(x,\tau)u
\end{equation}
The basic diffusion problem can be solved using a kernel solution:
\begin{equation}
K(x,y;t)=\int_{\Omega}d\mathfrak{m}(\lambda).e^{-E_{\lambda}t}\Phi_{\lambda}^{*}(x)\Phi_{\lambda}(y)+\sum_{n=1}^{n_{max}}|C_{n}|^{2}e^{-E_{n}t}\Psi_{n}^{*}(x)\Psi_{n}(y)
\end{equation}
Briefly, the moving parts of this expression are the continuous eigenfunction \(\Phi_{\lambda}(y)\), the discrete eigenfunctions \(\Psi_{n}(y)\), the cutoff for the discrete states \(n_{max}\), which may be some finite number or infinity for a purely discrete system, and the measures for the continuous spectrum \(\mathfrak{m}(\lambda)\) and discrete states \(|C_{n}|^{2}\).
Applying the method of separation of variables, we can extract the eigenfunctions by solving the time independent differential equations:

   \begin{equation*}
       u(x,\tau)=\phi(x)\Upsilon(\tau)
   \end{equation*} 
   \begin{equation}
       -\dfrac{1}{\Upsilon(\tau)}\dfrac{d\Upsilon(\tau)}{d\tau}=\dfrac{1}{\phi(x)}\left(\dfrac{\sigma^{2}}{2}\dfrac{d^{2}\phi}{dx^{2}}+\mu\dfrac{d\phi}{dx}+V(x,\tau)\phi(x)\right)
   \end{equation}
Assuming a time-independent potential/killing function \(V(x,\tau)=V(x)\), it is straightforward to write down the eigenfunction solution for the PDE system, which enables us to determine the kernel. 
\begin{equation}
    \dfrac{\sigma^{2}}{2}\dfrac{d^{2}\phi}{dx^{2}}+\mu\dfrac{d\phi}{dx}+V(x)\phi(x)=E\phi(x)
\end{equation}
\begin{equation}
    \dfrac{d\Upsilon(\tau)}{d\tau}=-E\Upsilon(\tau)
\end{equation}
It is clear that to understand the nature of the solution space, we need to determine the spectrum of the eigenvalues via the constant \(E\).

\section{Completeness of Eigenfunction Representations}
\label{sec:3}
The fundamental notion which we shall now exploit relates to the completeness of basis sets, and the relationship with orthonormal eigenfunctions. The spectral theory of the kernel states that we can find a set of eigenfunctions with the completeness property expressed through the delta function expansion:
\begin{equation}
    \delta(x-y)=\int_{\Omega}d\mathfrak{m}(\lambda).\Phi_{\lambda}^{*}(x)\Phi_{\lambda}(y)+\sum_{n=1}^{n_{max}}|C_{n}|^{2}\Psi_{n}^{*}(x)\Psi_{n}(y)
\end{equation}
The integral in this case is over the continuous eigenvalue; the sum over the discrete set of states. If we are able to determine the delta function expansion, we can find the kernel in a similar way. If we succeed in this, we will have solved the diffusion problem, as we may then construct the kernel solution which encodes all the measurable information in the system. 

The two parts which compose the delta function are the continuous and discrete parts of the state. The continuous state, defined through the eigenfunction \(\Phi_{\lambda}(y)\), has measure \(\mathfrak{m}(\lambda)\). In particular, many of the situations considered in this paper have a single continuous eigenfunction defined in a way which is analogous to a scattered state in quantum physics. In the same way that the Born equation gives the solution to the quantum equations for a particle scattered to infinity, the continuous state in our situation will often have an integral measure which is integrated over the positive part of the plane or section thereof. The second part of the completeness relationship is the discrete states; the upper bound \(n_{max}\) in this case takes into account the generalised example whereby there may be a finite rather than infinite set of orthogonal polynomials that span the discrete set of eigenfunctions. 

\section{Potential Theory}
\label{sec:4}
If we consider contracts which have positive prices, it is clear that if the process is positive valued, then the average is by necessity a positive valued function. By a similar logic, the potential or killing function \(V(x,\tau)\) must necessarily have values defined in the positive plane. Two examples are given in the potential function sketched below:


\begin{figure}[hp]
\centering

\includegraphics[scale=.60]{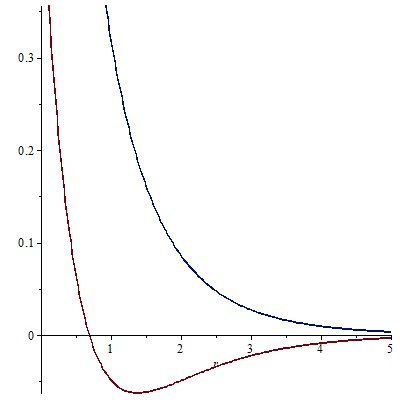}
\captionsetup{justification=centering}
\caption{Potential functions for the positive plane}
\end{figure}

The upper (blue) line in this case represents a potential which generates a continuous eigenstate. A particle released at any point on this curve will travel to infinity. The lower (red) line gives a different situation. Particles released above the x-axis will travel to infinity, but a particle released below the axis will oscillate. The spectrum in the first case will be a single continuous eigenvalue, and in the latter we will have a spectrum with both a continuous and a discrete part. In many ways we may understand these two different scenarios as analogous to scattered and bound eigenstates in quantum theory. 

We shall now show a simple technique that we can use to find simplified, transformed versions of the basic diffusion equations. The fundamental diffusion may be written:

\begin{equation}
    a(x)\dfrac{d^{2}\phi}{dx^{2}}+b(x)\dfrac{d\phi}{dx}+c(x)\phi(x)=0
\end{equation}
In terms of the drift and diffusion, we have:
\begin{equation}
    a(x)=\dfrac{\sigma^{2}(x)}{2},\thinspace b(x)=\mu(x),\thinspace c(x)=V(x)-E
\end{equation}
and the Bose invariant, defined through:
\begin{equation}
    \dfrac{d^{2}\hat{\phi}}{dx^{2}}+\mathcal{I}(x)\hat{\phi}(x)=0
\end{equation}
\begin{equation}
    \mathcal{I}(x)=\dfrac{4ac-b^{2}+2(ba'-ab')}{4a^{2}}
\end{equation}
The basic transformations of the eigenfunction are given by the speed and scale factors, defined through the fundamental symmetry relation:
\begin{equation}
    \left[a(x)\right]^{-1}h(x)\mathcal{H}(x,\partial_{x})\left[h(x)\right]^{-1}=\partial_{x}^{2}+\mathcal{I}(x)
\end{equation}
where the generator 
\begin{math}
    \mathcal{H}(x,\partial_{x})
\end{math}
is given by:
\begin{equation}
    \mathcal{H}(x,\partial_{x})=a(x)\dfrac{\partial^{2}}{\partial x^{2}}+b(x)\dfrac{\partial}{\partial x}+c(x)
\end{equation}
and the transformed eigenfunction is defined through the scale factor:
\begin{equation}
    \phi(x)=\dfrac{\hat{\phi}(x)}{h(x)}
\end{equation}
\begin{equation}
    h(x)=\exp\left(\int^{x}\dfrac{b(s)}{2a(s)}ds\right)
\end{equation}
Indeed, it is simple to see that the application of the Bose invariant to any diffusion problem results in a plain Gaussian diffusion, plus a transformed potential or killing function. This is a very powerful method as we shall show; the speed and scale measures of Feller 
\cite{feller1952parabolic} are obtained through:
\begin{equation}
    dm(x)=m(x)dx=\dfrac{e^{B(x)}}{a(x)}dx
\end{equation}
\begin{equation}
    s(x)=e^{-B(x)}
\end{equation}
\begin{equation}
    B(x)=\int^{x}\dfrac{b(s)}{a(s)}d\tau
\end{equation}
and, as we shall show, many of the integration measures for the diffusion processes which are important in this paper are related to these distributions. Finally, the killed measure is given by the formula:
\begin{equation}
    k(y)=\dfrac{c(y)e^{B(y)}}{4a(y)}
\end{equation}
Note that the killing measure is essentially just the killing potential multiplied by the scale factor. For an excellent discussion of these topics, the interested reader is directed to the works of Kuznetzov and Albanese \cite{albanese2007transformations}, also Sousa and Yakubovich \cite{sousa2017spectral} which contains an excellent discussion of the construction of kernels using the Titchmarsh-Kodaira theorem. Our argument shall differ from theirs in that we shall be examining the case where the continuous and discrete parts of the spectrum must both be taken into account.

\section{Asian Options Pricing Formulae}
\label{sec:5}
We shall report on some results that are available in the literature. Geman and Yor \cite{geman1993bessel} gave the canonical result for the fixed strike call option by exploiting the scaling property of Brownian motion via:
\begin{align}
    \mathbb{E}_{T}\left[\left(\dfrac{1}{T-t_{0}}\int_{t_{0}}^{T}S_{u}du-K\right)^{+}\right]
\end{align}
\begin{equation}
    =\dfrac{4S_{t}e^{-r(T-t)}}{(T-t_{0})\sigma^{2}}\mathbb{E}_{\sigma^{2}s/4}^{\mathbb{Q}}\left[\left(\int_{0}^{\sigma^{2}s/4}\exp\left(2W_{s'}+\dfrac{4}{\sigma^{2}}\nu s'\right)ds'-\dfrac{\mathcal{K}\sigma^{2}}{4}\right)^{+}\right]
\end{equation}

We may price call options for a fixed strike by using the transition probability density:

\begin{equation}
	C(T,0;S_{0},K)=\dfrac{4S_{0}e^{-rT}}{T\sigma^{2}}\int_{0}^{\infty}(u-\dfrac{\mathcal{K}\sigma^{2}}{4})^{+}f(u,0;\tau)du
\end{equation}
where the fundamental diffusion equation is defined through:
\begin{equation}
	2u^{2}\dfrac{\partial^{2}f}{\partial u^{2}}+\left[(2\zeta+1)u+1\right]\dfrac{\partial f}{\partial u}=-\dfrac{\partial f}{\partial t}
\end{equation}
and the variable \(u\) is the exponential Brownian motion:
\begin{equation}
	u=\int_{0}^{\sigma^{2}s/4}\exp\left(2\left[W_{s'}+\zeta s'\right]\right)ds'
\end{equation}
Other known results include the floating strike call option formula, developed in \cite{craddock2014integral}, which can be easily derived via the following means. The price of the option is defined through the payoff function:
\begin{equation}
	C_{f}(S_{T},A_{T},T)=\max\left(S_{T}-\dfrac{A_{T}}{T},0\right)
\end{equation}
from which it is possible to show that diffusion equation that relates to this system is given through the degenerate parabolic equation:
\begin{equation}
	\dfrac{\partial C_{f}}{\partial t}+\dfrac{\sigma^{2}S^{2}}{2}\dfrac{\partial^{2}C_{f}}{\partial S^{2}}+S\dfrac{\partial C_{f}}{\partial\xi}+rS\dfrac{\partial C_{f}}{\partial S}-rC_{f}=0
\end{equation}
To solve this expression, as in \cite{craddock2014integral}, one may employ the ansatz 
\begin{math}
	C_{f}(S,A,t)=Sf(z,t)
\end{math}
Again, we find the differential equation which defines the probability density:
\begin{equation}
-\dfrac{\partial f}{\partial t}+\dfrac{1}{2}\sigma^{2}z^{2}\dfrac{\partial^{2}f}{\partial z^{2}}-\left(\dfrac{1}{T}+rz\right)\dfrac{\partial f}{\partial z}=0
\end{equation}
The redundant variable in this case is defined through the time average in a similar way to the Yor system, via:
\begin{equation}
	\xi_{t}=\int_{0}^{t}S_{u}du
\end{equation}
The change of variables in this case is given by the relationship:
\begin{equation}
	z_{t}=\dfrac{1}{S}(k-\xi_{t})=\dfrac{1}{S}\left(k-\dfrac{A_{t}}{T}\right)
\end{equation}

\section{Application of the Bose Invariant}
\label{sec:6}
If we can solve either the system of Craddock \cite{craddock2014integral} or Yor \cite{geman1993bessel}, it is possible to find the solution of the other problem. We shall now show how to directly address the Yor problem using the Bose invariant. In an analogous way to the fixed strike, the scale factor for the Yor equation is readily shown to be inverse gamma distributed, via:
\begin{equation}
	h(u)=\exp\left(\int^{u}\dfrac{\left[(2\zeta+1)s+1\right]}{4s^{2}}ds\right)=e^{-1/(4u)}u^{1/4+\zeta/2}
\end{equation}
Evaluating the isomorphism of the generator of the diffusion, we find:
\begin{equation}
	\dfrac{1}{2u^{2}}e^{-1/(4u)}u^{1/4+\zeta/2}\hat{\mathcal{L}}\left(e^{1/(4u)}u^{-1/4-\zeta/2}f(u)\right)=\partial_{u}^{2}+\mathcal{I}(u)
\end{equation}
\begin{equation}
	2u^{2}\dfrac{\partial^{2}f}{\partial u^{2}}+\left[(2\zeta+1)u+1\right]\dfrac{\partial f}{\partial u}=Ef\Leftrightarrow\hat{\mathcal{L}}f=Ef
\end{equation}
We recover the Bose invariant potential, either by using the scale factor, or the formula using the PDE coefficients:
\begin{equation*}
\mathcal{I}(u)=\dfrac{1}{16u^{4}}+\dfrac{\left(-\dfrac{\zeta}{4}+\dfrac{3}{8}\right)}{u^{3}}+\dfrac{\left(-\dfrac{\zeta^{2}}{4}+\dfrac{\zeta}{4}-\dfrac{E}{2}+\dfrac{3}{16}\right)}{u^{2}}	
\end{equation*}
\begin{equation}
	=\dfrac{C_{1}}{u^{2}}+\dfrac{C_{2}}{u^{3}}+\dfrac{C_{3}}{u^{4}}
\end{equation}
The solution to the PDE system is then given by the Whittaker-type confluent hypergeometric function:
\begin{equation}
	\phi(u)=\dfrac{\hat{\phi}(u)}{h(u)}\sim Ce^{1/(4u)}u^{3/4-\zeta/2}W_{\alpha,ip}\left(\dfrac{1}{2u}\right)
\end{equation}
The scalar parameters in this expression are defined through:
\begin{equation}
	\alpha=\dfrac{3}{4}-\dfrac{\zeta}{2},\thinspace ip=\dfrac{1}{4}\sqrt{4(\zeta^{2}-\zeta)+8E+1}
\end{equation}
If we use the inverse variable 
\begin{math}
	x=(2u)^{-1}
\end{math}
we recover the Whittaker equation directly. Further transformations give the Morse potential, viz.:
\begin{equation}
	\dfrac{\partial^{2}F(x)}{\partial x^{2}}+\left(\dfrac{C_{1}}{x^{2}}+\dfrac{C_{2}}{x}+C_{3}\right)F(x)=0
\end{equation}
where the Whittaker-type solution may be understood through the spherical relation 
\begin{equation}
	F(x)=uF(u^{-1})
\end{equation}
. The Morse potential is given by the substitution 
\begin{math}
	u=e^{-r}
\end{math}
where the PDE system in the new variables is given by:
\begin{equation}
	\left(\partial_{r}^{2}+\mathcal{J}(r)\right)\Phi(r)=0
\end{equation}
\begin{equation}
	\mathcal{J}(r)=\dfrac{\gamma^{2}}{4}e^{-2r}-\dfrac{\nu}{2}e^{-r}+\dfrac{\alpha}{4}-\dfrac{1}{2}
\end{equation}
This is, indeed, the potential function described in the previous sections. For various different regimes of the parameters, we are able to tune between a purely scattering potential and a bound potential state. The spectrum is then given by a direct sum over the two contributions from the discrete and continuous parts of the state.

\section{Spectral Theory of Whittaker/Morse Potential}
\label{sec:7}
The potential derived in the previous section can be related to the spectral theory of the kernel. The standard form of the Whittaker equation can be written as:
\begin{equation}
	\left(-\partial_{z}^{2}+\dfrac{\gamma^{2}}{4}-\dfrac{\nu}{2z}+\left(\dfrac{\alpha}{4}-\dfrac{1}{4}\right)\dfrac{1}{z^{2}}\right)\phi(z)=0
\end{equation}
The potential function in this case is given by:
 \begin{equation}
	V(z)=\dfrac{\gamma^{2}}{4}-\dfrac{\nu}{2z}+\left(\dfrac{\alpha}{4}-\dfrac{1}{4}\right)\dfrac{1}{z^{2}}
\end{equation}
Note the generalised problem defined through the killed process:
 \begin{equation}
	\dfrac{\partial f}{\partial t}=\dfrac{\partial^{2}f}{\partial x^{2}}-V(x)f(x,t)
\end{equation}
As before, we have the expression for the kernel:
\begin{equation} K(x,y;t)=\int_{\Omega}d\mathfrak{m}(\lambda).e^{-E_{\lambda}t}\Phi_{\lambda}^{*}(x)\Phi_{\lambda}(y)+\sum_{n=1}^{n_{max}}|C_{n}|^{2}e^{-E_{n}t}\Psi_{n}^{*}(x)\Psi_{n}(y) 
\end{equation}
Now, it is known that the Green's function is given by the Laplace transform of the kernel via:
 \begin{equation}
	G(x,y;E)=\int_{0}^{\infty}e^{-Et}K(x,y;t)dt
\end{equation}
where, for the Whittaker potential, a known result from \cite{grosche1998handbook} gives the Green's function as:
\begin{equation*}        
	\pi^{2}\dfrac{\Gamma(1/2-\kappa+s)}{\Gamma(1+2s)}W_{\kappa,s}(x_{>})M_{\kappa,s}(x_{<})
\end{equation*}
\begin{equation}
	\resizebox{.80 \textwidth}{!}{ $
		=\int_{0}^{\infty}\dfrac{u\sinh(2\pi u)\Gamma\left(1/2-\kappa-iu\right)\Gamma\left(1/2-\kappa+iu\right)}{s^{2}+u^{2}}W_{\kappa,iu}(x)W_{\kappa,iu}(y)du 
		$}  
\end{equation}
\begin{equation}
	\resizebox{.80 \textwidth}{!}{ $
		+\pi^{2}e^{-(x+y)/2}\sum_{n=0}^{[Re(\kappa)-1/2]}\dfrac{\alpha n!}{\Gamma(\alpha+n+1)}\dfrac{(xy)^{(\alpha+1)/2}}{s^{2}-\alpha^{2}/4}L_{n}^{(\alpha)}(x)L_{n}^{(\alpha)}(y)
		$}
\end{equation}
The first part of this expression gives the continuous eigenstate, and the second part is therefore the discrete component. The eigenvalues are related by:
\begin{equation}
	\alpha=2\kappa-2n-1
\end{equation}
Concisely, the Green's function may be written as the compact product:
\begin{equation}
	G(x,y;E)=Ce^{-(x+y)/2}W_{\kappa,s}(x_{<})M_{\kappa,s}(x_{>})
\end{equation}

\section{Discrete Green's Function and Bessel Polynomials}
\label{sec:8}
We shall now discuss the discrete part of the Green's function as given by the Bessel polynomials \cite{olver2010digital}. These polynomials differ from the classical orthogonal polynomials, as for a fixed total energy, we have only a finite set of orthogonal polynomials in the positive plane. This is to be compared with e.g. the Hermite polynomials, which possess an infinite dimensional set of orthogonal eigenfunctions. The basic expression for the Bessel polynomials is given in \cite{grosswald1951some,krall1949new} as the Laguerre-type formula:
 \begin{equation}
	y_{n}(x;\alpha,\beta)=(-1)^{n}n!\left(\dfrac{x}{\beta}\right)^{n}L_{n}^{(1-2n-\alpha)}\left(\dfrac{\beta}{x}\right)
\end{equation}
One confluent form of the Bessel polynomials is given by:
\begin{equation}
	y_{n}(x;\alpha+1,1/2)=(-1)^{n}2^{n}x^{n}n!L_{n}^{(-2n-\alpha)}\left(\dfrac{1}{2x}\right)
\end{equation}
which satisfies the differential equation:
\begin{equation}
	2x^{2}\dfrac{d^{2}f}{dx^{2}}+\left(2(\alpha+1)x+1\right)\dfrac{df}{dx}-2\left[n(n+\alpha)\right]f=0
\end{equation}
Note that the eigenstates for this differential system are given through:
 \begin{equation}
	\Psi_{n}(x)=C_{n}^{\alpha}y_{n}(x;\alpha+1/2,1/2)
\end{equation}
and we see that the differential equation of interest can be easily related to the Asian option diffusion equation:
\begin{equation}
	2x^{2}\dfrac{d^{2}f}{dx^{2}}+\left((2\alpha+1)x+1\right)\dfrac{df}{dx}=2\left[n(n+\alpha)\right]f
\end{equation}

\section{Continuous Part of the Whittaker Kernel}
\label{sec:9}
If we examine the pricing diffusion for the Asian option, following the results in the previous sections, we have:
\begin{equation}
	2x^{2}\dfrac{d^{2}f}{dx^{2}}+\left((2\alpha+1)x+1\right)\dfrac{df}{dx}=Ef
\end{equation}
In this case, the energy eigenvalue is given through the relation:
\begin{equation}
	E=\dfrac{1}{2}\alpha(1-\alpha)-\dfrac{p^{2}}{2}-\dfrac{1}{8}
\end{equation}
with continuous eigenfunction solution:
\begin{equation}
	\Phi_{p}(x)=C(p)e^{1/(4x)}x^{\kappa}W_{\kappa,ip/2}\left(\dfrac{1}{2x}\right)
\end{equation}
\begin{equation}
	\kappa=-\alpha/2+3/4
\end{equation}
We shall now show how one may evaluate the measure of the space. The scale measure is given through the relationship \cite{borodin2015handbook}:
\begin{equation}
	dm(x)=m(x)dx=\dfrac{e^{B(x)}}{a(x)}dx
\end{equation}
\begin{equation}
	B(x)=\int^{x}\dfrac{b(s)}{a(s)}ds
\end{equation}
Evaluating for the diffusion equation, we find the inverse gamma distribution:
\begin{equation}
	m(x)=\dfrac{1}{2}e^{-1/(2x)}x^{-3/2+\alpha}
\end{equation}
Writing out the continuous part of the kernel, we find the expression:
  \begin{equation}
	K_{c}(x,y;\tau)=\int_{0}^{\infty}e^{-E(p)\tau}\Phi_{p}^{*}(x)\Phi_{p}(y)d\mathfrak{m}(p)
\end{equation}
Now, using the symmetries of the Whittaker function, it is not too hard to see that we have the relationship:
 \begin{equation}
	\Phi_{p}^{*}(x)=\Phi_{-p}(x)=\Phi_{p}(x)
\end{equation}
which may be rewritten for our eigenfunction as:
\begin{equation}
	W_{\kappa,ip}^{*}(z)=W_{\kappa,-ip}(z)=W_{\kappa,ip}(z)
\end{equation}
The principal aim is then to recover a completeness relationship of the type:
\begin{equation}
	K_{c}(x,y;0)=\lambda\delta(x-y)
\end{equation}
Following results found in \cite{szmytkowski2010orthogonality}, a known formula for orthonormality of Whittaker functions is given by the delta function relation:
\begin{equation}
	\int_{0}^{\infty}C_{\kappa,\mu}^{2}\dfrac{W_{\kappa,i\mu}(x)W_{\kappa,i\mu'}(x)}{x^{2}}dx=\delta(\mu-\mu')
\end{equation}
and the measure distribution is then given by:
\begin{equation}
	d\mathfrak{m}=C_{\kappa,\mu}^{2}\dfrac{dx}{x^{2}}=\dfrac{\mu\sinh(2\pi\mu)\Gamma\left(1/2-\kappa+i\mu\right)\Gamma\left(1/2-\kappa-i\mu\right)}{\pi^{2}x^{2}}
\end{equation}
Simplification using the properties of the gamma function then gives us the compact relation:
\begin{equation}
	d\mathfrak{m}(p)=\dfrac{1}{\pi^{2}x^{2}}p\sinh(2\pi p)|\Gamma\left(1/2-\kappa+ip\right)|^{2}dp
\end{equation}
from which we can evaluate the continuous part of the kernel as:
\begin{equation}
	K_{c}(x,y;\tau)=C_{\kappa}\int_{_{0}}^{\infty}e^{-p^{2}\tau/2}W_{\kappa,ip}\left(\dfrac{1}{2x}\right)W_{\kappa,ip}\left(\dfrac{1}{2y}\right)d\mathfrak{m}(p)
\end{equation}
Finally, by using inverse coordinate transforms, we find the simplified form of the kernel:
\begin{equation}
	\tilde{K}_{c}(x,y;\tau)=K_{c}((2x)^{-1},(2y)^{-1};\tau)
\end{equation}
\begin{equation}
	=\int_{_{0}}^{\infty}e^{-2p^{2}\tau}W_{\kappa,ip}\left(x\right)W_{\kappa,ip}\left(y\right)d\mathfrak{m}(p)
\end{equation}
These are the basic properties of the continuous spectral solution. We note that this is the important contribution in the parameter space defined by financial contracts. However, in contrast to say e.g. the results in \cite{sousa2017spectral}, we take the position that for full understanding of the kernel solution, both the discrete and continuous parts must be taken into account. We shall discuss in the following sections an extension of the Titchmarsh-Kodaira theorem according to this scenario and some ways in which the results of Sousa and Yakubovich may be modified to derive a similar result.

\section{Pricing Asian Options via Laguerre Interpolation}
\label{sec:10}
We shall now calculate Asian options prices using the kernel formulae developed through this type of analysis. The papers of Linetsky \cite{linetsky2004spectral}, Geman and Yor \cite{geman1993bessel}, and Sousa and Yakubovich \cite{sousa2017spectral} are of use in developing the necessary stochastic calculus to specify the pricing formula. For completeness, we shall briefly review the basic equations, and then show how one can use techniques from classical numerical analysis to construct various approximations for pricing this option. 
The basic analysis begins with the exponential Brownian motion:
\begin{equation}
    A_{\tau}^{(\nu)}=\int_{0}^{\tau}e^{2(B_{u}+\nu u)}du
\end{equation}
Put and call prices are computed using the time-invariant solution to the differential equation \cite{geman1993bessel, linetsky2004spectral}:
\begin{equation}
	P^{(\nu)}(k,\tau)=\mathbb{E}\left[(k-A_{\tau}^{(\nu)})^{+}\right]
\end{equation}
\begin{equation}
	V_{p}=e^{-rT}\mathbb{E}\left[(K-\mathcal{A}_{T})^{+}\right]=e^{-rT}\left(\dfrac{4S_{0}}{\sigma^{2}T}\right)P^{(\nu)}(k,\tau)
\end{equation}
\begin{equation}
	V_{c}=e^{-rT}\mathbb{E}\left[(\mathcal{A}_{T}-K)^{+}\right]=V_{p}+\dfrac{1-e^{-rT}}{rT}S_{0}-e^{-rT}K
\end{equation}
Using the results from the previous sections, it is possible to evaluate the price of the Asian option via:
\begin{equation}
	\resizebox{.85 \textwidth}{!}{ $
		P^{(\nu)}(k,\tau)=e^{-\nu^{2}\tau/2}\dfrac{(2k)^{-\kappa}e^{-1/(4k)}}{8\pi^{2}}\int_{0}^{\infty}e^{-p^{2}\tau/2}W_{\kappa,ip/2}\left(\dfrac{1}{2k}\right)d\mathfrak{m}+..
		$ } 
\end{equation}
where the measure is defined by:
\begin{equation}
	d\mathfrak{m}=\left|\Gamma\left(\dfrac{\nu+ip}{2}\right)\right|^{2}\sinh\pi p.pdp
\end{equation}
which we derived previously. Various integrals give interesting relationships between the relevant groups of special functions, viz.:
\begin{equation*}
	\resizebox{.93 \textwidth}{!}{ $
		\dfrac{1}{2\pi i}\int_{-i\infty}^{+i\infty}e^{\xi q}q^{-\nu/2}K_{2ip}\left(\sqrt{8qx}\right)dq=\dfrac{1}{\sqrt{8e^{x}}}\xi^{(\nu-1)/2}e^{-x/\xi}W_{(1-\nu)/2,ip/2}\left(\dfrac{2x}{\xi}\right)
		$ }
\end{equation*}
where the scalar parameters are defined through:
\begin{equation*}
	\nu=\dfrac{2r}{\sigma^{2}}-1, k=\dfrac{\tau K}{S_{0}}, \tau=\dfrac{T\sigma^{2}}{4}, \kappa=-\dfrac{(\nu+3)}{2}
\end{equation*}
We shall now show how these formulae may be computed using Laguerre quadrature of the integrals implied through the expectation value.
\begin{equation*}
	\resizebox{.90 \textwidth}{!}{ $
		P^{(\nu)}(k,\tau)=\int_{0}^{\infty}f(p)dp\approx\sum_{i=2}^{N-1}f(p_{i})\Delta p-\dfrac{1}{2}[f(p_{1})-f(p_{N})]\Delta p $ }
\end{equation*}
The integral implied through the option price can be approximated using the trapezoidal rule. However, for Laguerre interpolation we use the set of Lagrange points:
\begin{equation}
	\int_{a}^{b}f(x)W(x)dx=\sum_{i=1}^{n}w_{i}(x_{i})f(x_{i})
\end{equation}
where the weights give the integral-sum relations via:
\begin{equation}
	\int_{0}^{\infty}f(x)e^{-x}dx=\sum_{i=1}^{n}w_{i}f(x_{i})
\end{equation}
and are defined by the Laguerre points:
\begin{equation}
	w_{i}=\dfrac{1}{(n+1)^{2}[L_{n+1}(x_{i})]^{2}}
\end{equation}
In terms of the exponential geometric Brownian motion, by virtue of the positive nature of the GBM, we know that the average can only be a positive non-zero process. This makes Laguerre or generalised Laguerre quadrature the appropriate choice in this situation, because this is also defined over the positive part of the plane.  Other systems \cite{linetsky2004spectral} rely on a spectral decomposition over the zeroes of the eigenfunction.
\begin{equation}
	P_{b}^{(\nu)}(k,\tau)=\sum_{n=1}^{\infty}\dfrac{e^{-\lambda_{n,b}t}}{\left\langle \Psi,\Psi\right\rangle _{p_{n,b}}}\int_{0}^{k}(k-y)\Psi(y;\lambda_{n,b})\mathfrak{m}(y)dy
\end{equation}
There is the question of convergence of the spectral method especially in the low volatility/low drift case. We expect to recover the heat equation for some limit. This is particularly relevant to markets today in an economy experiencing low interest rates for a prolonged time. A more efficient way of establishing a numerical pricing scheme is by using quadrature of the integral:
\begin{equation}
	\int_{a}^{b}f(x)W(x)dx=\sum_{i=1}^{n}w_{i}(x_{i})f(x_{i})
\end{equation}
Gauss-Laguerre quadrature has the weighting function:
\begin{equation}
	\int_{0}^{\infty}f(x)e^{-x}dx=\sum_{i=1}^{n}w_{i}f(x_{i})
\end{equation}
Weights are readily calculated using the orthogonality property of Laguerre polynomials:
\begin{equation}
	w_{i}=\dfrac{1}{(n+1)^{2}[L_{n+1}(x_{i})]^{2}}
\end{equation}
To compute the integral, it is important to recognise that numerically stable forms of the confluent hypergeometric function are required.
\begin{equation*}
	\resizebox{.90 \textwidth}{!}{ $
		U(a,b,x)=\dfrac{\pi}{\sin\pi b}\left(\dfrac{M(a,b,x)}{\Gamma(1+a-b)\Gamma(b)}+x^{1-b}\dfrac{M(1+a-b,2-b,x)}{\Gamma(a)\Gamma(2-b)}\right)
		$}
\end{equation*}
\begin{equation}
	M(a,b,x)=e^{x}M(b-a,b,-x)
\end{equation}
\begin{equation}
	\tilde{W}_{\kappa,\mu}(z)=e^{-z/2}z^{b+1/2}\tilde{U}\left(b-a+1/2,1+2b,z\right)
\end{equation}
The parameters used in the integral are:
\begin{equation}
	z=\dfrac{1}{2k}, \kappa=-\dfrac{(\nu+3)}{2}, \mu=\dfrac{ip}{2}
\end{equation}
With these formulae for the Whittaker function, we are able to recover prices for Asian options in a variety of regimes. The final ingredients in this mathematical algorithm are the precision requirements of the integral. Using the concept of exponential addition and multiplication for extra precision, it was simple to control any numerical errors in the procedure:
\begin{equation}
	P^{(\nu)}(k,\tau)=\int_{0}^{\infty}g(p;\nu,\tau,k)dp
\end{equation}
\begin{equation}
	g(p;\nu,\tau,k)=\mathfrak{Re}\left(\dfrac{V_{4}}{8\pi^{2}}\exp\left(\ln V_{1}+\ln V_{2}+\ln V_{3}\right)\right)
\end{equation}
We were able to demonstrate convergence for a series of well-known test cases \cite{linetsky2004spectral}, however the low volatily case was shown to require extra care. The current R package (fAsianOptions, now defunct) fails on several test cases due to limitations in precision. The cases from \cite{linetsky2004spectral} are outlined in the table below.

\subsubsection{Parameters for Test Cases from \cite{linetsky2004spectral}}
\begin{center}
\begin{tabular}{|c|c|c|c|c|c|c|c|}
\hline 
Case & $r$ & $\sigma$ & $T$ & $S_{0}$ & $\nu$ & $\tau$ & $k$\tabularnewline
\hline 
\hline 
1 & $0.02$ & $0.10$ & $1$ & $2.0$ & $3$ & 0.0025 & $0.0025$\tabularnewline
\hline 
2 & $0.18$ & $0.30$ & $1$ & $2.0$ & $3$ & $0.0225$ & $0.0225$\tabularnewline
\hline 
3 & $0.0125$ & $0.25$ & $2$ & $2.0$ & $-0.6$ & $0.03125$ & $0.03125$\tabularnewline
\hline 
4 & $0.05$ & $0.50$ & $1$ & $1.9$ & $-0.6$ & $0.0625$ & $0.065789$\tabularnewline
\hline 
5 & $0.05$ & $0.50$ & $1$ & $2.0$ & $-0.6$ & $0.0625$ & $0.625$\tabularnewline
\hline 
6 & $0.05$ & $0.50$ & $1$ & $2.1$ & $-0.6$ & $0.0625$ & $0.059524$\tabularnewline
\hline 
7 & $0.05$ & $0.50$ & $2$ & $2.0$ & $-0.6$ & $0.125$ & $0.125$\tabularnewline
\hline 
\end{tabular}
\par\end{center}

Note that these figures are by no means an exhaustive search over
the parameter space. The final three columns, given by the parameters
$\nu,\tau,k$ are derived from the first set of parameters, which
define the stochastic system. It isn't intuitively clear what distinguishes
the first case as being difficult and less likely to converge, other
than the combined situation of both low drift, as expressed through
the parameter $r$, as well as a smaller value for the volatility,
as given by $\sigma$. The following tables compare the results for various methods of calculating Asian option prices.

\subsubsection{Table 1- Asian Option Call Prices}
\begin{center}
\begin{tabular}{|c|c|c|c|c|}
\hline 
Case & Spectral Method & Trapezoidal Rule & Vecer Method & Monte Carlo\tabularnewline
\hline 
\hline 
1 & $14.7694$ & $14.7694$ & $0.0559843109$ & $0.055929(24)$\tabularnewline
\hline 
2 & $0.2183875465$ & $0.2183875465$ & $0.2183869858$ & $0.21704(83)$\tabularnewline
\hline 
3 & $0.1722687410$ & $0.1722687383$ & $0.1722682714$ & $0.173279(89)$\tabularnewline
\hline 
4 & $0.193173790$ & $0.193145214$ & $0.1931716834$ & $0.1922(11)$\tabularnewline
\hline 
5 & $0.246415690$ & $0.246397641$ & $0.2464144912$ & $0.2450(13)$\tabularnewline
\hline 
6 & $0.306220364$ & $0.306208936$ & $0.3062195980$ & $0.3055(14)$\tabularnewline
\hline 
7 & $0.3500952189$ & $0.348084942$ & $0.3498467545$ & $0.3496(20)$\tabularnewline
\hline 
\end{tabular}
\par\end{center}

\subsubsection{Asian Options Prices via Laguerre Quadrature}
\begin{center}
\begin{tabular}{|c|c|c|}
\hline 
 & Laguerre & \tabularnewline
\hline 
Case & Put & Call\tabularnewline
\hline 
\hline 
1 & 9.55{*}10\textasciicircum 5 & 9.55{*}10\textasciicircum 5\tabularnewline
\hline 
1{*}(200 points) & 0.0362615331589084254661 & 0.0559968558698675989430\tabularnewline
\hline 
2 & 0.0585969850989125223683 & 0.2183875465955682429392\tabularnewline
\hline 
3 & 0.1476815247399004291038 & 0.1722687384166216016910\tabularnewline
\hline 
4 & 0.2423229661816367419105 & 0.1931459861530763782535\tabularnewline
\hline 
5 & 0.1980339582371130057155 & 0.2463981292071246238756\tabularnewline
\hline 
6 & 0.1603039232499463874824 & 0.3062092452185299874597\tabularnewline
\hline 
7 & 0.2545623438519754839573 & 0.3481391470608648743438\tabularnewline
\hline 
\end{tabular}
\par\end{center}
As we can see, with the exception of the pathological case 1, the Laguerre method is highly accurate. This, while not particularly useful for quantitative purposes, may be of relevance to physical situations where the concept of exponential Brownian motion is of interest. To the author's knowledge, these are the most accurate figures produced for these types of options prices and form a useful sandbox test set for validating other models.

The low volatility case is specified by the parameter set:
\begin{equation}
	r=0.02,\sigma=0.1,T=1,K=2.0
\end{equation}
\begin{equation}
	S_{0}=2.0,\nu=3,\tau=0.0025,k=0.0025
\end{equation}
Understanding why this case is difficult when the values seem reasonable is the challenge in deriving formulae for Asian options pricing. The low volatility/low drift approximation of equation can be expected to have some quirks due to limiting behaviour of Whittaker function. We ran some simulations to compare this method against standard techniques, inc. Monte Carlo, spectral theory, discretised PDE, other integral approximations. As the data shows, it seems to hold up well across a variety of situations- with the caveat of an error control term. We can define our convergence region via the difference between a Monte Carlo sim with a known tolerance and the output from our formula. Numerical experiments allow us to probe this region for different parameter sets and to find upper bounds on the number of abscissa points in the quadrature. The chart below shows this convergence for the pathological case as discussed in detail in the previous sections.

\begin{figure}[hp]
\centering
\includegraphics[scale=.15]{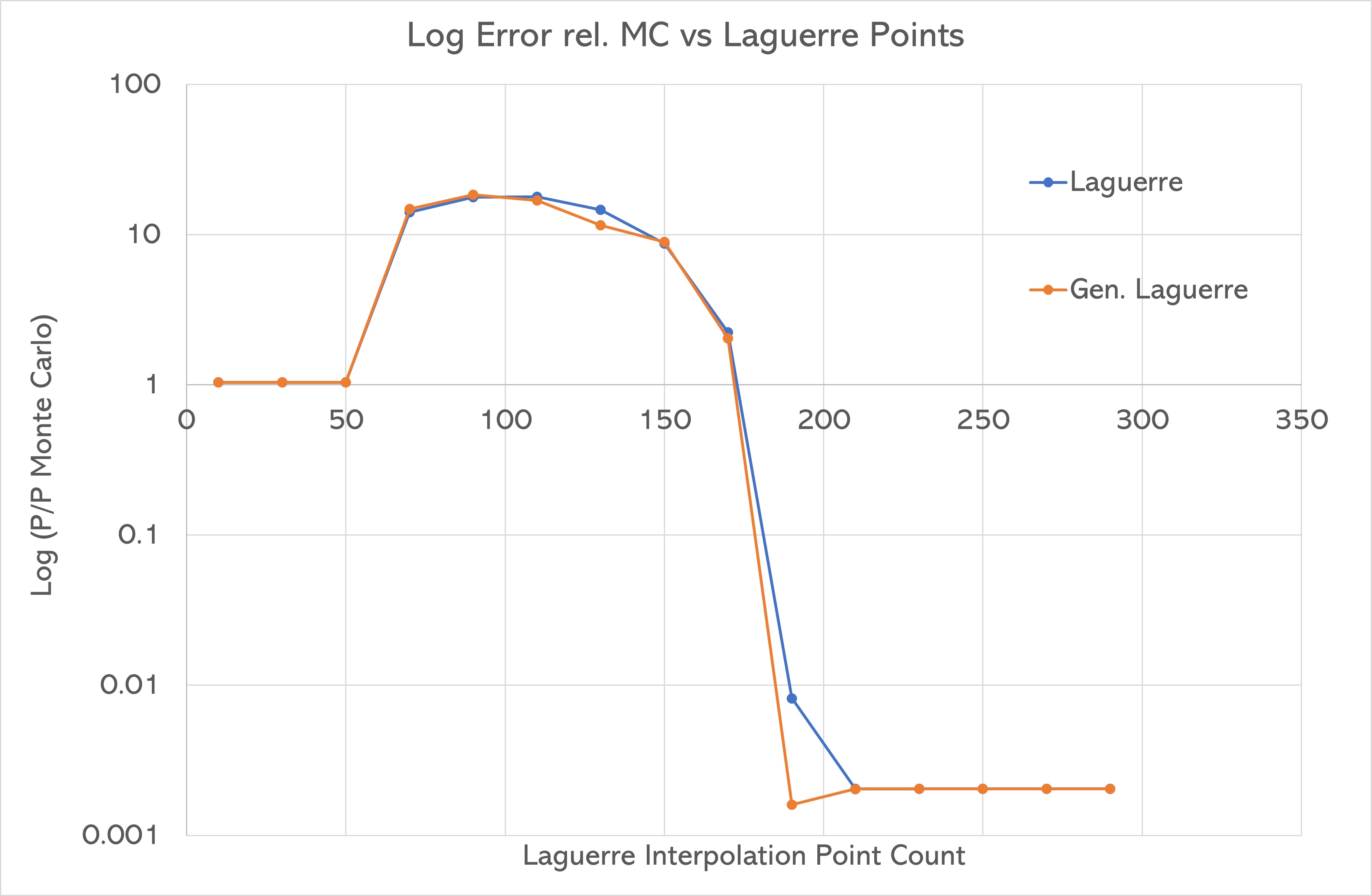}

\captionsetup{justification=centering}
\caption{Convergence for Case 1- Low volatility, low drift}
\end{figure}

\section{Spectral Expansions}
\label{sec:12}

We shall now discuss some generalities related to the calculation in the previous sections. Sousa and Yakubovich \cite{sousa2017spectral} give the expression for the Titchmarsh-Kodaira theorem via the integral transform pair:
\begin{equation}
	(\mathcal{F}f)_{i}(\lambda)=\tilde{f}_{i}(\lambda)=\int_{a}^{b}f(x)w_{i}^{*}(x,\lambda)r(x)dx
\end{equation} 
\begin{equation}
	(\mathcal{F}^{-1}\tilde{f})(x)=\int_{0^{-}}^{\infty}\sum_{i,j=1}^{2}\tilde{f}_{i}(\lambda)w_{j}(x,\lambda)d\rho_{ij}(\lambda)
\end{equation}
The measure in the inverse transform is given by the Titchmarsh-Kodaira theorem:
\begin{equation}
	\rho_{ij}(\lambda_{1},\lambda_{2})=\lim_{\delta\rightarrow0}\lim_{\epsilon\rightarrow0}\dfrac{1}{2\pi i}\int_{\lambda_{1}+\delta}^{\lambda_{2}-\delta}\left[m_{ij}^{+}(\lambda+i\epsilon)-m_{ij}^{+}(\lambda-i\epsilon)\right]d\lambda
\end{equation}
\begin{equation}
	=\lim_{\delta\rightarrow0}\lim_{\epsilon\rightarrow0}\dfrac{1}{2\pi i}\int_{\lambda_{1}+\delta}^{\lambda_{2}-\delta}\left[m_{ij}^{-}(\lambda+i\epsilon)-m_{ij}^{-}(\lambda-i\epsilon)\right]d\lambda
\end{equation}
where the kernel of the system is defined through the eigenfunction expansion:
\begin{equation}
	K(x,y;\lambda)=\sum_{i,j=1}^{2}m_{ij}^{-}(\lambda)w_{i}(x,\lambda)w_{j}(y,\lambda^{*})
\end{equation}
where \begin{math}
	x<y
\end{math}

\begin{equation}
	K(x,y;\lambda)=\sum_{i,j=1}^{2}m_{ij}^{+}(\lambda)w_{i}(x,\lambda)w_{j}^{*}(y,\lambda^{*})
\end{equation}
analogously, where \begin{math}
	x>y
\end{math}
These different representations give the complex spherical and hyperbolic decompositions for the kernel. The functions
\begin{math}
	w_{i}(x,\lambda)
\end{math}
are naturally then associated with the eigenfunctions of the differential equations which define the system, via the solutions of the equation for the kernel \begin{math}
	\mathcal{L}u=\lambda u
\end{math}. From this definition one can go on to generate various representations for the Green's function, resolvent and other such operators relevant to the study of such differential equations.

\subsection{Heat Kernel for Hermite Polynomials}

We shall now briefly comment on the natural generalisation of this type of system. Note that this is only strictly true for a system with a single continuous degree of freedom. As in the previous examples we have examined, a simple system is given by the Kontorovich-Lebedev transform, where the continuous eigenvalue represents a continuum of scattered states. However, in the case where we consider the heat equation equivalent of the harmonic oscillator, it is simple to see that the relevant diffusion equation is defined through:
\begin{equation}
	-\dfrac{\partial u}{\partial t}=-\dfrac{1}{2}\dfrac{\partial^{2}u}{\partial x^{2}}+\dfrac{1}{2}x^{2}u
\end{equation}
hence the eigenvalue equation is given by:
\begin{equation}
	E\phi(x)=-\dfrac{1}{2}\dfrac{d^{2}\phi}{dx^{2}}+\dfrac{1}{2}x^{2}\phi
\end{equation}
with solution:
\begin{equation}
	\phi(x)=C_{1}\phi_{1}(x)+C_{2}\phi_{2}(x)
\end{equation}
\begin{equation}
	\phi_{1}(x)=\dfrac{M_{E/2,1/4}(x^{2})}{\sqrt{x}}
\end{equation}
\begin{equation}
    =-\dfrac{1}{4}\dfrac{\Gamma\left(1/4-E/2\right)}{\sqrt{\pi}2^{E/2-1/4}}x^{3/2}\left(D_{E-1/2}(\sqrt{2}x)-D_{E-1/2}(-\sqrt{2}x)\right)
\end{equation}
\begin{equation}
	\phi_{2}(x)=\dfrac{W_{E/2,1/4}(x^{2})}{\sqrt{x}}=\dfrac{H_{E-1/2}(x)}{2^{E-1/2}}e^{-x^{2}/2}=\dfrac{D_{E-1/2}(\sqrt{2}x)}{2^{E/2-1/4}}
\end{equation}
If we consider known results from quantum mechanics \cite{feynman2010quantum}, it is clear that the physical solution is defined through the function
\begin{math}
	\phi_{2}(x)
\end{math}; to quantise the system in the appropriate fashion we must therefore have 
\begin{math}
	H_{E-1/2}(x)=H_{n}(x)
\end{math}, 
hence the energy eigenstates are defined by:
\begin{equation}
	E_{n}=n+\dfrac{1}{2}
\end{equation}
which, up to scaling, gives the correct equipartition law for blackbody radiation. If we then calculate the nature of the first solution, we find it must take the form:
\begin{equation}
	\phi_{1}(x)=-\dfrac{1}{4}\dfrac{\Gamma\left(-n/2\right)}{\sqrt{\pi}2^{n/2}}\left(D_{n}(\sqrt{2}x)-D_{n}(-\sqrt{2}x)\right)
\end{equation}
and it is obvious that this represents a solution that is scattered in a forwards and backwards direction in space. Returning to the physical solution, the Mehler kernel is given by the infinite sum:
\begin{equation}
	K(x,y;t)=\sum_{n=0}^{\infty}e^{-E_{n}t}\psi_{n}(x)\psi_{n}^{*}(y)
\end{equation}
whence, upon simplification
\begin{equation}
	K(x,y;t)=\sum_{n=0}^{\infty}e^{-(n+1/2)t}|c_{n}|^{2}H_{n}(x)H_{n}(y)e^{-(x^{2}+y^{2})/2}
\end{equation}
The normalisation of the eigenfunctions is defined through:
\begin{equation}
	\int_{0}^{\infty}|\psi_{n}(x)|^{2}dx=|c_{n}|^{2}\int_{0}^{\infty}H_{n}^{2}(x)e^{-x^{2}}dx=1
\end{equation}
\begin{equation}
	\int_{0}^{\infty}H_{n}^{2}(x)e^{-x^{2}}dx=\sqrt{\pi}2^{n}n!
\end{equation}
\begin{equation}
	|c_{n}|^{2}=\dfrac{1}{\sqrt{\pi}2^{n}n!}
\end{equation}
Writing out the kernel, we find:
\begin{equation*}
	K(x,y;t)=\sum_{n=0}^{\infty}\dfrac{e^{-(n+1/2)t}}{\sqrt{\pi}2^{n}n!}H_{n}(x)H_{n}(y)e^{-(x^{2}+y^{2})/2}
\end{equation*}
\begin{equation}
    =\dfrac{e^{-t/2}}{\sqrt{\pi}}\sum_{n=0}^{\infty}\dfrac{(e^{-t}/2)^{n}}{n!}H_{n}(x)H_{n}(y)e^{-(x^{2}+y^{2})/2}
\end{equation}
\begin{equation}
	\sum_{n=0}^{\infty}\dfrac{(\rho/2)^{n}}{n!}H_{n}(x)H_{n}(y)e^{-(x^{2}+y^{2})/2}=\dfrac{1}{\sqrt{1-\rho^{2}}}\exp\left(\dfrac{4xy\rho-(1+\rho^{2})(x^{2}+y^{2})}{2(1-\rho^{2})}\right)
\end{equation}
Using \begin{math}
	\rho=e^{-t}
\end{math}, the kernel is found via the concise formula:
\begin{equation}
	K(x,y;t)=\dfrac{e^{-t/2}}{\sqrt{\pi}}\dfrac{1}{\sqrt{1-e^{-2t}}}\exp\left(\dfrac{4xye^{-t}-(1+e^{-2t})(x^{2}+y^{2})}{2(1-e^{-2t})}\right)
\end{equation}
\begin{equation}
	=\dfrac{1}{\sqrt{2\pi\sinh(t)}}\exp\left(xy\mathrm{cosech}(t)-(x^{2}+y^{2})\coth(t)/2\right)
\end{equation}
This is a known result and demonstrates that using appropriate mathematical logic that one can derive compact, succinct formulae for kernels by using the eigenfunction expansions. We shall now show how this relates to the Bessel polynomials as considered earlier in the paper.

\subsection{Bessel Polynomials as Transforms and Kernels}

Now, let us consider the fundamental differences here. In the example for the Asian option pricing equation, as we have shown, there is a discrete and a continuous component, both of which contribute to the spectral solution given through the kernel. The discrete component in this case has only a finite number of energy states, and we also have a continuous eigenstate which will contribute. The natural generalisation of the Titchmarsh-Kodaira formula will then be given by:
\begin{equation}
	K(x,y;t)=\sum_{i=0}^{n_{max}}e^{-E_{i}t}|c_{i}|^{2}\Phi_{i}(x,\lambda)\Phi_{i}^{*}(y,\lambda)+\int_{\Omega}e^{-E_{\lambda}t}\Psi_{\lambda}(x)\Psi_{\lambda}^{*}(y)d\mathfrak{m}(\lambda)
\end{equation}
In this case we actually have two solutions to the eigenvalue equation 
\begin{math}
	\mathcal{L}u=Eu
\end{math}
depending on the values of the parameters which define the differential equation. It is interesting to consider the nature of the integral transform pairs that by necessity underlay the Bessel polynomials, which are the discrete part complementary to the Whittaker transform.

\begin{equation}
	(\mathcal{F}f)_{i}(\lambda)=\tilde{f}_{i}(\lambda)=\int_{a}^{b}f(x)w_{i}^{*}(x,\lambda)r(x)dx
\end{equation}
We shall now briefly discuss the Bessel polynomials and other finite sets of orthogonal functions. For any infinite set of orthogonal polynomials, we may write the following:
\begin{equation}
	f(x)=\sum_{n=0}^{\infty}A_{n}p_{n}(x)
\end{equation}
Assume we have orthogonality defined over some interval, then we may extract the coefficients using the orthogonality property:
\begin{equation}
	A_{n}=\int_{a}^{b}f(x)p_{n}(x)dx
\end{equation}
where
\begin{equation}
	\int_{a}^{b}p_{n}(x)p_{n'}(x)dx=\delta_{nn'}
\end{equation}
Now if we consider orthogonal polynomials of this type, we indeed have the transform pair:
\begin{equation}
	A_{n}=\int_{a}^{b}f(x)p_{n}(x)dx
\end{equation}
\begin{equation}
	f(x)=\sum_{n=0}^{\infty}A_{n}p_{n}(x)
\end{equation}
Although there is an orthogonality implied in the complex plane, we shall not consider this as it is beyond the scope of this simple discussion. Using results from \cite{hamza1974properties}, the following is known for the generalised Bessel polynomials 
\begin{math}
	y_{n}(x;a,b)
\end{math} (notation ours) with support in
\begin{math}
	[0^{+},\infty)
\end{math}:
\begin{equation}
	\int_{0}^{\infty}\left(\dfrac{x}{b}\right)^{1-a}e^{-x/b}y_{m}(x;a,b)y_{n}(x;a,b)dx=b(n!)\Gamma(2-a-n)\delta_{mn}
\end{equation}
This is the appropriate form for use in the following. The weight function is then:
\begin{equation}
	w_{b}(x)=\left(\dfrac{x}{b}\right)^{1-a}e^{-x/b}
\end{equation}
which we recognise as being gamma distributed in
\begin{math}
	x
\end{math}, and inverse gamma in 
\begin{math}
	b
\end{math}. Obviously this case differs significantly from the situation with the standard classical orthogonal polynomial series. Let us therefore expand a function defined in the positive plane into the Bessel polynomials via:
\begin{equation}
	f(x)=\sum_{n=0}^{\infty}A_{n}y_{n}(x;a,b)
\end{equation}
\begin{equation}
	A_{n}=\dfrac{1}{b(n!)\Gamma(2-a-n)}\int_{0}^{\infty}f(x)y_{n}(x;a,b)dx
\end{equation}
We shall call this the Bessel polynomial transform pair. Interestingly, there is a result from \cite{pintoux2010direct} which is of relevance in determining the correct expansion. The authors of the series of papers \cite{pintoux2011dothan, pintoux2010direct} report the following expansion equation:
\begin{equation*}
	x^{-p}=\dfrac{2^{-p}}{2\pi^{2}}\int_{0}^{\infty}u\sinh(\pi u)\left|\Gamma\left(-\dfrac{p}{2}+\dfrac{iu}{2}\right)\right|^{2}K_{iu}(x)du
\end{equation*}
\begin{equation}
    +2^{-p+1}\sum_{k=0}^{\left\lfloor p/2\right\rfloor }\dfrac{(p-2k)!}{k!(p-k)!}K_{p-2k}(x)
\end{equation}
which is true for all \begin{math}
	p
\end{math}. We can see that in this expansion that the first part is given by a Kontorovich-Lebedev transformation, and the second part is given by a Laguerre-type summation. It is this second part that is equivalent to the Bessel polynomial transform as discussed above. Indeed, using the definition of the Bessel polynomial \cite{olver2010digital}, we may write:
\begin{equation}
	y_{n}(x)=\sqrt{\dfrac{2}{\pi x}}e^{1/x}K_{n+1/2}(x^{-1})
\end{equation}
\begin{equation}
	K_{p-2k}(x)=\sqrt{\dfrac{\pi}{2x}}e^{-1/x}y_{p-2k-1/2}(x^{-1})
\end{equation}
The finite, discrete polynomial part of the transform is then given by the Bessel polynomial sum:
\begin{equation}
	\hat{\mathcal{T}}_{disc.}f(x)=2^{-p+1}\sum_{k=0}^{\left\lfloor p/2\right\rfloor }\dfrac{(p-2k)!}{k!(p-k)!}\sqrt{\dfrac{\pi}{2x}}e^{-x}y_{p-2k-1/2}(x^{-1})f(x,u)
\end{equation}
\begin{equation}
	=2^{-p+1}\sqrt{\dfrac{\pi}{2x}}e^{-x}\sum_{k=0}^{\left\lfloor p/2\right\rfloor }\dfrac{(p-2k)!}{k!(p-k)!}y_{p-2k-1/2}(x^{-1})f(x,u)
\end{equation}
The continuous part of the same transform implied through the formula is, of course, the generalised Kontorovich-Lebedev transform:
\begin{equation}
	\hat{\mathcal{T}}_{cont.}f(x,u)=\dfrac{2^{-p}}{2\pi^{2}}\int_{0}^{\infty}u\sinh(\pi u)\left|\Gamma\left(-\dfrac{p}{2}+\dfrac{iu}{2}\right)\right|^{2}K_{iu}(x)f(x,u)du
\end{equation}
as required. This completes our discussion of the spectral theory of the Asian option pricing problem. It is clear that the nature of the solutions to the modified Bessel equation requires us, for certain ranges of parameters, to take into account both the discrete and continuous parts which contribute to the solution as a whole. It is easily seen that there exist certain polynomial subspaces of finite dimensions both for this equation, and for the Whittaker equation more generally. In the latter situation, we know that the Bessel polynomials are of importance, however, it seems to have been overlooked that this is necessary in order that the completeness relationship be well defined and analytical in the case of the Bessel equation for all possible values of the parameters.

\section{Index Integration}

One simple technique that can be used to determine formulae for index
transformations are the Whittaker-Bessel-Legendre integrals, which
allow the reduction of more complicated integrals to known types involving
Legendre equations. We shall briefly show several integrals in which
this method of calculation is effective, and indicate some further
types of problems that may arise for more complicated scenarios that
may be of interest. 

\subsection{A Mixed Index Integral}

For example, if we take the completeness relationship:

\begin{equation}
\pi\delta(x-y)=\int_{0}^{\infty}p\sinh(\pi p)|\Gamma(-\mu+1/2+ip)|^{2}\mathcal{P}_{ip-1/2}^{\mu}(x)\mathcal{P}_{ip-1/2}^{\mu}(y)dp
\end{equation}

it is possible to show, using some results from integral theory that the following addition formula is available:
\begin{equation*}
\dfrac{2\Gamma(1-\mu)}{\pi(a+b)}W_{\mu,1/2}(2(a+b))
\end{equation*}
\begin{equation}
    =\dfrac{2}{\pi^{2}}\int_{0}^{\infty}p\sinh(\pi p)|\Gamma(-\mu+1/2+ip)|^{2}\dfrac{W_{\mu,ip}(2a)}{a}\dfrac{W_{\mu,ip}(2b)}{b}dp
\end{equation}

We shall now show how another different type of index integral may
be found using the technique of completeness. We shall ``mix'' over
the basis sets defined by the modified Bessel and Mehler-Fock functions
viz.:
\begin{equation}
\mathit{I}=\int_{0}^{\infty}p\sinh(\pi p)\left|\Gamma\left(\dfrac{\mu}{2}+ip\right)\right|^{2}K_{ip}(a)\mathcal{P}_{ip-1/2}^{\mu}(y)dp
\end{equation}

Using known integral relations between the Macdonald and Mehler-Fock
functions, we may write:
\begin{equation}
\int_{1}^{\infty}e^{-ax}(x^{2}-1)^{-\mu/2}\mathcal{P}_{ip-1/2}^{\mu}(x)dx=\sqrt{\dfrac{2}{\pi}}a^{\mu-1/2}K_{ip}(a)
\end{equation}

or:
\begin{equation}
K_{ip}(a)=\sqrt{\dfrac{\pi}{2}}a^{-\mu+1/2}\int_{1}^{\infty}e^{-ax}(x^{2}-1)^{-\mu/2}\mathcal{P}_{ip-1/2}^{\mu}(x)dx
\end{equation}

hence the integral rolls over into the nested formula:
\begin{equation}
\mathit{I}=\sqrt{\dfrac{\pi}{2}}a^{-\mu+1/2}\int_{0}^{\infty}p\sinh(\pi p)\left|\Gamma\left(\dfrac{\mu}{2}+ip\right)\right|^{2}\mathcal{P}_{ip-1/2}^{\mu}(y)dp
\end{equation}
\begin{equation}
\times\int_{1}^{\infty}e^{-ax}(x^{2}-1)^{-\mu/2}\mathcal{P}_{ip-1/2}^{\mu}(x).dx
\end{equation}

Invoking the Fubini theorem and interchanging the order of integrals,
we find:
\begin{equation}
\mathit{I}=\sqrt{\dfrac{\pi}{2}}a^{-\mu+1/2}\int_{1}^{\infty}dx.e^{-ax}(x^{2}-1)^{-\mu/2}
\end{equation}
\begin{equation}
\times\int_{0}^{\infty}p\sinh(\pi p)\left|\Gamma\left(\dfrac{\mu}{2}+ip\right)\right|^{2}\mathcal{P}_{ip-1/2}^{\mu}(y)\mathcal{P}_{ip-1/2}^{\mu}(x)dp
\end{equation}
\begin{equation}
=\sqrt{\dfrac{\pi}{2}}a^{-\mu+1/2}\int_{1}^{\infty}e^{-ax}(x^{2}-1)^{-\mu/2}\pi\delta(x-y)dx
\end{equation}
\begin{equation}
=\sqrt{\dfrac{\pi^{3}}{2}}a^{-\mu+1/2}e^{-ay}(y^{2}-1)^{-\mu/2}
\end{equation}

The result then states that 
\begin{equation}
\int_{0}^{\infty}p\sinh(\pi p)\left|\Gamma\left(\dfrac{\mu}{2}+ip\right)\right|^{2}K_{ip}(a)\mathcal{P}_{ip-1/2}^{\mu}(y)dp
\end{equation}
\begin{equation}
=\sqrt{\dfrac{\pi^{3}}{2}}a^{-\mu+1/2}e^{-ay}(y^{2}-1)^{-\mu/2}
\end{equation}

This simple technique of expansion, interchange and resolution using
the completeness identity has stark and profound results for a large
class of index transforms. As we shall show, many different integrals
can be evaluated by this method.

\subsection{Second Mixed Integral}

We shall now show how to treat another type of mixed integral of a
similar kind. In this case we have e.g.:

\begin{equation}
I=\int_{0}^{\infty}p\sinh(\pi p)\left|\Gamma\left(\dfrac{1}{2}-\mu+\dfrac{ip}{2}\right)\right|^{2}W_{\mu,ip}(2x)K_{ip}(y)dp
\end{equation}
\begin{equation}
=\sqrt{\dfrac{\pi}{2}}xy^{-\mu+1/2}\int_{0}^{\infty}p\sinh(\pi p)\left|\Gamma\left(\dfrac{1}{2}-\mu+\dfrac{ip}{2}\right)\right|^{2}\sqrt{\dfrac{2}{\pi}}y^{\mu-1/2}\dfrac{W_{\mu,ip}(2x)}{x}K_{ip}(y)dp
\end{equation}
\begin{equation}
=\sqrt{\dfrac{\pi}{2}}xy^{-\mu+1/2}\int_{0}^{\infty}dp.p\sinh(\pi p)\left|\Gamma\left(\dfrac{1}{2}-\mu+\dfrac{ip}{2}\right)\right|^{2}
\end{equation}
\begin{equation}
\times\int_{1}^{\infty}e^{-xu}\left(\dfrac{u+1}{u-1}\right)^{-\mu/2}\mathcal{P}_{ip-1/2}^{\mu}(u)du\int_{1}^{\infty}e^{-yv}(v^{2}-1)^{-\mu/2}\mathcal{P}_{ip-1/2}^{\mu}(v)dv
\end{equation}

It is important at this point to take the correct symmetries for the
Whittaker, modified Bessel and Mehler-Fock functions. Indeed, as they
are all spherical functions, we may extend the support of the function
to take in negative values, via.:
\begin{equation}
\mathcal{P}_{ip-1/2}^{\mu}(u)=\mathcal{P}_{ip-1/2}^{\mu}(-u)
\end{equation}
\begin{equation}
W_{\mu,ip}(2x)=W_{\mu,ip}(-2x)
\end{equation}
\begin{equation}
K_{ip}(y)=K_{ip}(-y)
\end{equation}
The integral is invariant if we take these definitions, hence we must
ensure that the final answer respects the same symmetry.
\begin{equation}
I=\int_{0}^{\infty}p\sinh(\pi p)\left|\Gamma\left(\dfrac{1}{2}-\mu+\dfrac{ip}{2}\right)\right|^{2}W_{\mu,ip}(2x)K_{ip}(y)dp
\end{equation}

Rearrangement via the Fubini-Tonelli theorem yields:
\begin{equation}
I=\sqrt{\dfrac{\pi}{2}}\pi.xy^{-\mu+1/2}\int_{1}^{\infty}\int_{1}^{\infty}dudv.e^{-(xu+yv)}\left(\dfrac{u+1}{u-1}\right)^{-\mu/2}(v^{2}-1)^{-\mu/2}
\end{equation}
\begin{equation}
\times\int_{0}^{\infty}dp.p\sinh(\pi p)\left|\Gamma\left(\dfrac{1}{2}-\mu+\dfrac{ip}{2}\right)\right|^{2}\mathcal{P}_{ip-1/2}^{\mu}(u)\mathcal{P}_{ip-1/2}^{\mu}(v)
\end{equation}
\begin{equation}
=\sqrt{\dfrac{\pi^{3}}{2}}xy^{-\mu+1/2}\int_{1}^{\infty}\int_{1}^{\infty}dudv.e^{-(xu+yv)}\left(\dfrac{u+1}{u-1}\right)^{-\mu/2}(v^{2}-1)^{-\mu/2}\delta(u-v)
\end{equation}
\begin{equation}
=\sqrt{\dfrac{\pi^{3}}{2}}xy^{-\mu+1/2}\int_{1}^{\infty}du.e^{-(x+y)u}\left(\dfrac{u+1}{u-1}\right)^{-\mu/2}(u^{2}-1)^{-\mu/2}
\end{equation}
\begin{equation}
=\sqrt{\dfrac{\pi^{3}}{2}}xy^{-\mu+1/2}\int_{1}^{\infty}du.e^{-(x+y)u}\left(\dfrac{u+1}{u-1}\right)^{-\mu/2}(u-1)^{-\mu/2}(u+1)^{-\mu/2}
\end{equation}
\begin{equation}
=\sqrt{\dfrac{\pi^{3}}{2}}xy^{-\mu+1/2}\int_{1}^{\infty}du.e^{-(x+y)u}(u+1)^{-\mu}
\end{equation}
One way in which to understand this is through the following integral \cite{gradshteyn2014table} 3.382.2:
\begin{equation}
\int_{u}^{\infty}(x-u)^{\nu}e^{-\mu x}dx=\mu^{-\nu-1}e^{-u\mu}\Gamma(\nu+1)
\end{equation}
\begin{equation}
\int_{1}^{\infty}(x-1)^{-\nu}e^{-\mu x}dx=\mu^{\nu-1}e^{-\mu}\Gamma(-\nu+1)=\int_{\infty}^{1}(-1)^{-\nu}(1+u)^{-\nu}e^{\mu u}du
\end{equation}
\begin{equation}
=-(-1)^{-\nu}\int_{1}^{\infty}(1+u)^{-\nu}e^{\mu u}du
\end{equation}
\begin{equation}
\int_{1}^{\infty}(1+u)^{-\nu}e^{\mu u}du=-(-1)^{\nu}\mu^{\nu-1}e^{-\mu}\Gamma(-\nu+1)
\end{equation}

The phase factor is therefore cancelled, and we arrive at:
\begin{equation}
\int_{0}^{\infty}p\sinh(\pi p)\left|\Gamma\left(\dfrac{1}{2}-\mu+\dfrac{ip}{2}\right)\right|^{2}W_{\mu,ip}(2x)K_{ip}(y)dp
\end{equation}
\begin{equation}
=\sqrt{\dfrac{\pi^{3}}{2}}xy^{-\mu+1/2}e^{-(x+y)}(x+y)^{\mu-1}\Gamma(1-\mu,-2(x+y))
\end{equation}

In the sector where the parameters are defined through:
\begin{equation}
\mu\leq1/2
\end{equation}

the formula simplifies, yielding that of Oberhettinger \cite{oberhettinger1961tables} pp17, 34
\begin{equation}
\int_{0}^{\infty}p\sinh(\pi p)\left|\Gamma\left(\dfrac{1}{2}-\mu+\dfrac{ip}{2}\right)\right|^{2}W_{\mu,ip}(2x)K_{ip}(y)dp
\end{equation}
\begin{equation}
=\sqrt{\dfrac{\pi^{3}}{2}}xy^{-\mu+1/2}e^{-(x+y)}(x+y)^{\mu-1}\Gamma(1-\mu)
\end{equation}

This is a perfect application for complex analysis and residue calculus
and shall form the basis of a future paper. 

\subsection{Cosine Transform}

Asian options may be treated in a similar fashion by expanding the
Whittaker function into the Legendre state. If we take the formula
derived for Asian option pricing and perform this operation, we find:
\begin{equation}
\int_{0}^{\infty}e^{-p^{2}\tau}p\sinh(\pi p)\left|\Gamma\left(\dfrac{1}{2}-\mu+\dfrac{ip}{2}\right)\right|^{2}W_{\mu,ip}(2x)dp
\end{equation}
\begin{equation}
=x\int_{0}^{\infty}e^{-p^{2}\tau}p\sinh(\pi p)\left|\Gamma\left(\dfrac{1}{2}-\mu+\dfrac{ip}{2}\right)\right|^{2}\dfrac{W_{\mu,ip}(2x)}{x}dp
\end{equation}
\begin{equation}
=x\int_{0}^{\infty}dp.e^{-p^{2}\tau}p\sinh(\pi p)\left|\Gamma\left(\dfrac{1}{2}-\mu+\dfrac{ip}{2}\right)\right|^{2}
\end{equation}
\begin{equation}
    \times{\int_{1}^{\infty}e^{-xu}\left(\dfrac{u+1}{u-1}\right)^{-\mu/2}\mathcal{P}_{ip-1/2}^{\mu}(u)du}
\end{equation}
\begin{equation}
=x\int_{1}^{\infty}du.e^{-xu}\left(\dfrac{u+1}{u-1}\right)^{-\mu/2}
\end{equation}
\begin{equation}
    \times{\int_{0}^{\infty}dp.e^{-p^{2}\tau}p\sinh(\pi p)\left|\Gamma\left(\dfrac{1}{2}-\mu+\dfrac{ip}{2}\right)\right|^{2}\mathcal{P}_{ip-1/2}^{\mu}(u)}
\end{equation}
Note that we have used the multiplication symbol here to indicate a nested integral in the interests of brevity. The inner integral here is of primary interest. We note the connection
between integrals of this type and the so-called Yor integrals. We
shall now show some methods that can be used to evaluate this integral
by relying on various transforms between groups of special functions.
Calculating the cosine transform of the Whittaker function:
\begin{equation}
g_{c}(y)=\int_{0}^{\infty}f(x)\cos(xy)dx
\end{equation}
\begin{equation}
g_{c}(y)=\int_{0}^{\infty}W_{\mu,ip}(2u)\cos(py)dp
\end{equation}
\begin{equation}
=\int_{0}^{\infty}u\int_{1}^{\infty}e^{-ux}\left(\dfrac{x+1}{x-1}\right)^{\mu/2}\mathcal{P}_{ip-1/2}^{\mu}(x)\cos(py).dxdp
\end{equation}
\begin{equation}
=u\int_{1}^{\infty}dx.e^{-ux}\left(\dfrac{x+1}{x-1}\right)^{\mu/2}\int_{0}^{\infty}dp.\mathcal{P}_{ip-1/2}^{\mu}(x)\cos(py)
\end{equation}

Oberhettinger \cite{oberhettinger2012tables} 12.85 gives the following cosine
transform of the Mehler-Fock function:
\begin{equation}
\int_{0}^{\infty}dp.\mathcal{P}_{ip-1/2}^{\mu}(x)\cos(py)=\sqrt{\dfrac{\pi}{2}}\dfrac{\left(x^{2}-1\right)^{\mu/2}\left(x-\cosh y\right)^{-\mu-1/2}}{\Gamma\left(1/2-\mu\right)}
\end{equation}
\begin{equation}
\cosh y<a
\end{equation}

so we obtain the cosine transform of the Whittaker function via:
\begin{equation}
g_{c}(y)=\sqrt{\dfrac{\pi}{2}}\dfrac{u}{\Gamma\left(1/2-\mu\right)}\int_{1}^{\infty}dx.e^{-ux}\left(x+1\right)^{\mu}\left(x-\cosh y\right)^{-\mu-1/2}
\end{equation}

Using formula from \cite{gradshteyn2014table} 3.385.3 
\begin{equation}
\int_{\alpha}^{\infty}dx.e^{-ax}\left(x+\beta\right)^{2\nu-1}\left(x-\alpha\right)^{2\rho-1}
\end{equation}
\begin{equation}
=\dfrac{(\alpha+\beta)^{\nu+\rho-1}}{a^{\nu+\rho}}\exp\left(\dfrac{(\beta-\alpha)a}{2}\right)\Gamma(2\rho)W_{\nu-\rho,\nu+\rho-1/2}\left([\alpha+\beta]a\right)
\end{equation}

This formula is not quite correct for our application. In the region
we need to have the following definition:
\begin{equation}
\int_{\alpha}^{\infty}dx.e^{-ax}\left(x+\alpha\right)^{2\rho-1}\left(x-\beta\right)^{2\nu-1}
\end{equation}
\begin{equation}
=\dfrac{(\alpha+\beta)^{\nu+\rho-1}}{a^{\nu+\rho}}\exp\left(\dfrac{(\beta-\alpha)a}{2}\right)\Gamma(2\nu)W_{\nu-\rho,\nu+\rho-1/2}\left([\alpha+\beta]a\right)
\end{equation}

Making the substitutions 
\begin{equation}
\int_{\alpha}^{\infty}dx.e^{-ax}\left(x+\alpha\right)^{2\rho-1}\left(x-\beta\right)^{2\nu-1}
\end{equation}
\begin{equation}
=\dfrac{(\alpha+\beta)^{\nu+\rho-1}}{a^{\nu+\rho}}\exp\left(\dfrac{(\beta-\alpha)a}{2}\right)\Gamma(2\nu)W_{\nu-\rho,\nu+\rho-1/2}\left([\alpha+\beta]a\right)
\end{equation}
\begin{equation}
\alpha=1,\beta=-\cosh y,2\rho-1=\mu,a=u,2\nu-1=-\mu-1/2
\end{equation}
\begin{equation}
\rho=\dfrac{\mu+1}{2}
\end{equation}
\begin{equation}
\nu=\dfrac{1}{2}\left(-\mu+1/2\right)=-\dfrac{\mu}{2}+\dfrac{1}{4}
\end{equation}
\begin{equation}
\nu-\rho=-\dfrac{\mu}{2}+\dfrac{1}{4}-\dfrac{\mu}{2}-\dfrac{1}{2}=-\mu-\dfrac{1}{4}
\end{equation}
\begin{equation}
\nu+\rho-1/2=-\dfrac{\mu}{2}+\dfrac{1}{4}+\dfrac{\mu+1}{2}-1/2=\dfrac{1}{4}
\end{equation}
\begin{equation}
\nu+\rho=3/4,\nu+\rho-1=-1/4
\end{equation}
\begin{equation}
\int_{1}^{\infty}dx.e^{-ux}\left(x+1\right)^{\mu}\left(x-\cosh y\right)^{-\mu-1/2}
\end{equation}
\begin{equation}
=\dfrac{(1+\cosh y)^{-1/4}}{u^{3/4}}\exp\left(\dfrac{-(\cosh y-1)u}{2}\right)\Gamma(1/2-\mu)W_{-\mu-1/4,1/4}\left([1+\cosh y]u\right)
\end{equation}
Using the hyperbolic identities and some simple trigonometry, it is
simple to reduce this integral further to:
\begin{equation}
\cosh y=\cosh\left(\dfrac{y}{2}+\dfrac{y}{2}\right)=\cosh^{2}\dfrac{y}{2}+\sinh^{2}\dfrac{y}{2}
\end{equation}
\begin{equation}
=2\cosh^{2}\dfrac{y}{2}-1=1+2\sinh^{2}\dfrac{y}{2}
\end{equation}
\begin{equation}
1+\cosh y=2\cosh^{2}\dfrac{y}{2}
\end{equation}
\begin{equation}
1-\cosh y=-2\sinh^{2}\dfrac{y}{2}
\end{equation}
\begin{equation}
\int_{1}^{\infty}dx.e^{-ux}\left(x+1\right)^{\mu}\left(x-\cosh y\right)^{-\mu-1/2}
\end{equation}
\begin{equation}
=\dfrac{(\cosh^{2}\dfrac{y}{2})^{-1/4}}{u^{3/4}}\exp\left(-u\sinh^{2}\dfrac{y}{2}\right)\Gamma(1/2-\mu)W_{-\mu-1/4,1/4}\left(u\cosh^{2}\dfrac{y}{2}\right)
\end{equation}

Using identities for the Whittaker function and parabolic cylinder
functions, we find:
\begin{equation}
W_{-\mu-1/4,1/4}\left(x\right)=W_{\mu+1/4,1/4}\left(x\right)=D_{2\mu}\left(\sqrt{2x}\right)\dfrac{x^{1/4}}{2^{\mu}}
\end{equation}
\begin{equation}
W_{-\mu-1/4,1/4}\left(u\cosh^{2}\dfrac{y}{2}\right)=D_{2\mu}\left(\sqrt{2u}\cosh\dfrac{y}{2}\right)\dfrac{u^{1/4}\cosh\dfrac{y}{2}}{2^{\mu}}
\end{equation}
\begin{equation}
\int_{1}^{\infty}dx.e^{-ux}\left(x+1\right)^{\mu}\left(x-\cosh y\right)^{-\mu-1/2}
\end{equation}
\begin{equation}
    =\dfrac{(\cosh^{2}\dfrac{y}{2})^{-1/4}}{u^{3/4}}\exp\left(-u\sinh^{2}\dfrac{y}{2}\right)\Gamma(1/2-\mu)D_{2\mu}\left(\cosh\dfrac{y}{2}\sqrt{2u}\right)\dfrac{(u\cosh^{2}\dfrac{y}{2}){}^{1/4}}{2^{\mu}}
\end{equation}
\begin{equation}
=\dfrac{1}{2^{\mu}u^{1/2}}\exp\left(-u\sinh^{2}\dfrac{y}{2}\right)\Gamma(1/2-\mu)D_{2\mu}\left(\cosh\dfrac{y}{2}\sqrt{2u}\right)
\end{equation}

Finalising the result for the index transform, we obtain:
\begin{equation}
g_{c}(y)=\sqrt{\dfrac{\pi}{2}}\dfrac{u}{\Gamma\left(1/2-\mu\right)}\int_{1}^{\infty}dx.e^{-ux}\left(x+1\right)^{\mu}\left(x-\cosh y\right)^{-\mu-1/2}
\end{equation}
\begin{equation}
=\sqrt{\dfrac{\pi}{2}}\dfrac{\Gamma(1/2-\mu)}{2^{\mu}u^{1/2}}\dfrac{u}{\Gamma\left(1/2-\mu\right)}\exp\left(-u\sinh^{2}\dfrac{y}{2}\right)D_{2\mu}\left(\cosh\dfrac{y}{2}\sqrt{2u}\right)
\end{equation}
\begin{equation}
=\sqrt{\dfrac{\pi u}{2}}2^{-\mu}\exp\left(-u\sinh^{2}\dfrac{y}{2}\right)D_{2\mu}\left(\cosh\dfrac{y}{2}\sqrt{2u}\right)
\end{equation}

which is consistent with a tabulated result of Oberhettinger using
an unknown method \cite{oberhettinger1961tables} 22.28. We have thus established,
with the caveat of the admission of the representation for the shifted
confluent hypergeometric function, the following cosine transform:
\begin{equation*}
\int_{0}^{\infty}W_{\mu,ip}(2u)\cos(py)dp
\end{equation*}
\begin{equation}
    =\sqrt{\dfrac{\pi u}{2}}2^{-\mu}\exp\left(-u\sinh^{2}\dfrac{y}{2}\right)D_{2\mu}\left(\cosh\dfrac{y}{2}\sqrt{2u}\right)=\mathscr{J}_{\mu}(y,u)
\end{equation}

\section{Conclusions and Future Directions}
\label{sec:13}
This paper has demonstrated a number of techniques that can be used to compute Asian options pricing. In particular, we have shown that the method of Gauss-Laguerre quadrature is a highly accurate algorithm that produces answers that are valid to a degree not possible to achieve via known methods, including the spectral method. Difficulties in convergence are easily overcome by increasing the number of abscissa points, which is not simple to achieve using other techniques. 

We have also shown that the problem of exponential Brownian motion is deeply associated with various groups of special functions, including those of Bessel, Mehler-Fock and Whittaker. By using spectral decompositions of the kernel over the continuous and discrete eigenfunctions, a large number of formulae related to various integral transforms are readily derived. This is a fruitful area of research that has relevance to the fields of quantitative finance as well as physical implications.

There are some further topics
of interest that present themselves that may be amenable to resolution
in a similar way to the calculations we have discussed. 
We shall outline some advanced techniques related to Whittaker functions that are of interest in studying these types of spectral theories and pricing algorithms. 
This study has skipped over the dynamics associated with the discrete part of the state. It is possible to show that it contributes via a similar integral, a part given by:
\begin{equation}
	\resizebox{.85 \textwidth}{!}{ $
		\sum_{n=0}^{[\nu/2]}e^{-2n(|\nu|-n)\tau}\dfrac{(-1)^{n}2(|\nu|-2n)}{\Gamma(1+|\nu|-n)}(2y)^{n-1-|\nu|}e^{-1/(2y)}L_{n}^{(|\nu|-2n)}\left(\dfrac{1}{2y}\right)
		$}
\end{equation}
The origin of this term is the poles of the Whittaker W-function. It is important to realise that the Green's function can be defined using the Cauchy integral formula, so we cannot neglect the discrete components contributing at the poles in the complex plane. We can look at both this expression and the continuous counterpart as being equivalent to the Yor integral which gives the pdf for the particle originating at zero. Research in the integral theory of special functions has turned up the following relationship:
\begin{equation}
	\int_{0}^{\infty}\nu\sinh(\pi\nu)\Gamma(\lambda+i\nu)\Gamma(\lambda-i\nu)K_{i\nu}(a)K_{i\nu}(b)d\nu
\end{equation}
\begin{equation}
	=\dfrac{\pi^{3/2}\Gamma(\lambda+1/2)}{2}\left(\dfrac{a+b}{2ab}\right)^{-\lambda}K_{\lambda}(a+b)
\end{equation}
There are relationships between the Whittaker systems and other groups of hyperbolic functions, i.e. associated Legendre, modified Bessel functions. It is an active topic of research aimed at better understanding the interaction between eigenstates, spectral theory and special functions.
\begin{acknowledgement}
The author wishes to acknowledge the support of the Department of Maths and Physical Sciences, UTS in conducting this research.
\end{acknowledgement}

\bigskip
%
\bibliographystyle{spmpsci}
\bibliography{references.bib}

\end{document}